\documentclass[journal,twoside]{IEEEtran}
\normalsize
\ifCLASSINFOpdf
\else
\fi
\usepackage[OT1]{fontenc}
\usepackage{float}
\usepackage{flushend}
\usepackage{graphicx}
\usepackage{tabularx}
\usepackage{booktabs}
\usepackage{multirow}
\usepackage{makecell}
\usepackage{colortbl}
\usepackage{CJK}
\usepackage{amsmath,epsfig,amssymb,bm,dsfont}
\usepackage{epstopdf}
\usepackage{longtable}
\usepackage{url}
\usepackage{pifont}
\usepackage{threeparttable}
\usepackage[linesnumbered,lined,ruled,vlined]{algorithm2e}
\usepackage{algpseudocode}
\usepackage{pgfplots}
\usepackage{mathtools}
\usepackage{verbatim}
\usepackage{bbding}
\usepackage{tikz}

\pgfplotsset{compat=1.14}

\ifCLASSOPTIONcompsoc
\usepackage[caption=false,font=normalsize,labelfont=sf,textfont=sf]{subfig}
\else
\usepackage[caption=false,font=footnotesize]{subfig}
\fi

\definecolor{mygray}{gray}{.9}

\newtheorem{Rem}{Remark}

\begin{document}
\title{Automatic Hybrid-Precision Quantization for MIMO Detectors}

\author{Yingmeng~Ge,
	Zhenhao~Ji,
	Yongming~Huang,~\IEEEmembership{Senior~Member,~IEEE},
	Zaichen~Zhang,~\IEEEmembership{Senior~Member,~IEEE},
	Xiaohu~You,~\IEEEmembership{Fellow,~IEEE},
	and~Chuan~Zhang,~\IEEEmembership{Senior~Member,~IEEE}%
\thanks{Y.~Ge, Z.~Ji, Y.~Huang, Z.~Zhang, X.~You, and C.~Zhang are with the LEADS of Southeast University, the National Mobile Communications Research Laboratory of Southeast University, and the Purple Mountain Laboratories, Nanjing 211189, China. \textit{(Yingmeng~Ge and Zhenhao~Ji contributed equally to this work.)} \textit{(Corresponding authors: Zaichen~Zhang and Chuan~Zhang.)}}}

\markboth{}
{Ge \MakeLowercase{\textit{et al.}}: Automatic Hybrid-Precision Quantization for MIMO Detectors}

\maketitle

\begin{abstract}
	In the design of wireless systems, quantization plays a critical role in hardware, which directly affects both area efficiency and energy efficiency. Being an enabling technique, the wide applications of multiple-input multiple-output (MIMO) heavily relies on efficient implementations balancing both performance and complexity. However, most of the existing detectors uniformly quantize all variables, resulting in high redundancy and low flexibility. Requiring both expertise and efforts, an in-depth tailored quantization usually asks for prohibitive costs and is not considered by conventional MIMO detectors.
In this paper, a general framework named the automatic hybrid-precision quantization (AHPQ) is proposed with two parts: integral quantization determined by probability density function (PDF), and fractional quantization by deep reinforcement learning (DRL).
Being automatic, AHPQ demonstrates high efficiency in figuring out good quantizations for a set of algorithmic parameters.
For the approximate message passing (AMP) detector, AHPQ achieves up to $58.7\%$ lower average bitwidth than the unified quantization (UQ) one with almost no performance sacrifice. The feasibility of AHPQ has been verified by implementation with $65$ nm CMOS technology. Compared with its UQ counterpart, AHPQ exhibits $2.97\times$ higher throughput-to-area ratio (TAR) with $19.3\%$ lower energy dissipation. Moreover, by node compression and strength reduction, the AHPQ detector outperforms the state-of-the-art (SOA) in both throughput ($17.92$\,Gb/s) and energy efficiency ($7.93$\,pJ/b). The proposed AHPQ framework is also applicable for other digital signal processing algorithms.

\end{abstract}


\section{Introduction}
\IEEEPARstart{W}{ith} the increasing demands of wireless communications, multiple-input multiple-output (MIMO) has gained wide attentions due to its high spectral efficiency (SE) and energy efficiency (EE) \cite{marzetta2010noncooperative}. To alleviate the unbearable complexity of the optimal MIMO detectors, \emph{Bayesian} message passing (BMP) algorithms have been widely considered to better balance performance and complexity, including belief propagation (BP)
, channel hardening-exploiting message passing (CHEMP)
, approximate message passing (AMP)
, expectation propagation (EP)
, and so on.
To bridge the gap between algorithms and implementations, hardware of those algorithms have been explored. In \cite{peng2012design} Peng \textit{et al.} presented a pipelined high-throughput ASIC implementation of BP detector. Implementations of message passing detectors (MPDs) based on CHEMP were given in \cite{tang20160, tang20210}. Those MPDs were further improved by deep neural network (DNN) towards higher efficiency and flexibility \cite{tan2019improving}. \cite{tan2019efficient, tan2020approximate} successfully reduced the complexity of EP detectors \cite{tang20181} with \emph{Neumann}-series approximations.


Although the aforementioned detectors raised the area and energy efficiency at the expense of performance, the main focus was on the arithmetic implementations but not on the quantization optimizations. As the very first and important step from algorithms (floating-point) towards implementations (fixed-point), the central issue of quantization optimization is to achieve the best trade-off between performance and complexity. Usually, the complexity is proportional to quantization length, but there is no such linearity in the relationship of performance and quantization. Although shorter quantization will penalize the performance, increasing quantization is not always cost-effective considering the \emph{performance saturation}. In addition, for each iteration of BMP detection, different variables have different requirements on quantization precision (intra-iteration). Those requirements also vary in different iterations (inter-iteration). Therefore, instead of figuring out an exactly-matching quantization, a unified quantization (UQ) for all variables, which is determined by limited numerical results and expertise, is usually adopted for implementation feasibility.

Saving design efforts, such UQ for all variables suffers from bitwidth redundancy and hardware overhead. Hence, hybrid-precision quantization with low design cost is highly expected by MIMO detectors. According to \cite{zhang2020artificial}, this optimization can be categorized as a \emph{problem difficult to model}, and AI techniques are expected to be promising helpers. Since the inter-iteration quantization can be transferred to the intra-iteration quantization by unfolding \cite{xu2020deep}, only the latter will be discussed here.


In this paper, the deep reinforcement learning (DRL) is utilized for automatic quantization. Though there are works on DRL-based MIMO detections \cite{jeon2018reinforcement, mo2021deep}, they mainly focus on algorithms. It is true that DRL-based quantization has been investigated for convolutional neural networks (CNNs) \cite{wang2019haq,elthakeb2020releq,lou2020autoq}, its application for MIMO detection is quite different for the following reasons.


\begin{enumerate}
	\item The structure of CNN is regular, mainly composed of convolutional layers, pooling layers, and fully connected layers. However, the MIMO detectors are diverse and without a universal structure. Taking BMP detectors as an example, EP, BP, and AMP are based on different factor graphs. Thus different BMP detectors vary in structures, bringing much more challenges.
	\item The accuracy requirement of MIMO detectors is much more demanding than that of CNNs. For example, the top-$5$ error rate of the $16$-bit ResNet-$50$ is $7.82\%$ \cite{lou2020autoq}. However, the required bit-error-rate (BER) of MIMO detector is usually lower than $10^{-3}$. Therefore, MIMO detectors are more sensitive to quantization noise.
	\item The aforementioned DRL-based automatic quantization frameworks for CNNs are all layer-wise (or kernel-wise). However, for MIMO detectors all variables require different quantizations to maximally squeeze the hardware.
\end{enumerate}

This paper devotes itself in proposing an automatic hybrid-precision quantization (AHPQ) for MIMO detectors. The contributions of this paper are listed as follows.
\begin{enumerate}
	\item For complexity consideration, the AHPQ works in two parts: integral part quantization (IPQ) based on probability density function (PDF), and fractional part quantization (FPQ) based on DRL. The general solution of IPQ is obtained by leveraging Monte Carlo simulations. For DRL-based FPQ, definitions of state, action, and reward are given. Two approaches addressing the reward misjudgment are proposed for better performance.
	\item A specific case, namely the quantization of AMP with nearest-neighbor approximation (NNA-AMP), is analyzed. The impact of the number of extracted variables and influence range of action is discussed with simulations. Compared with UQ, AHPQ presents a much lower quantization bitwidth.
	\item To validate the advantages of AHPQ from a hardware perspective, a hardware-friendly NNA-AMP (HF-AMP) detector by node compression (NC) and strength reduction is proposed. The VLSI architecture is given as well.
	\item The ASIC implementation of HF-AMP is presented.
The HF-AMP using AHPQ presents advantages in both area and energy efficiency compared with that using UQ. Moreover, HF-AMP using AHPQ outperforms the state-of-the-art (SOA) in both throughput ($17.92$\,Gb/s) and energy efficiency ($7.93$\,pJ/b).
\end{enumerate}

The reminder of this paper is organized as follows. In Section \ref{sec: Preliminaries}, the preliminaries of MIMO, DRL, and quantization are reviewed. In Section \ref{sec: DRLQ}, AHPQ is proposed with details. An application example of MIMO detectors is provided in Section \ref{sec: Application Examples}. In Section \ref{sec: HF AMP and VLSI}, the HF-AMP and its efficient VLSI architecture are presented. The ASIC implementations are given and compared in Section \ref{sec: Hardware Verification}. Section \ref{sec: Conclusion} concludes the entire paper.

\textbf{Notations:} $\mathbf{I}_n$ denotes the $n\times n$ identity matrix. $\mathbf{X}^{\mathsf{T}}$ is the transpose operation of the matrix $\mathbf{X}$. $\mathcal{N}\!\left(\boldsymbol{\mu}, \boldsymbol{\Sigma}\right)$ denotes the multi-variate \emph{Gaussian} distribution with mean vector $\boldsymbol{\mu}$ and covariance matrix $\boldsymbol{\Sigma}$, while $\mathcal{CN}\!\left(\boldsymbol{\mu}, \boldsymbol{\Sigma}\right)$ denotes the complex multi-variate \emph{Gaussian} distribution with complex mean vector $\boldsymbol{\mu}$ and complex covariance matrix $\boldsymbol{\Sigma}$. Operation ${\rm clip}(v, v_{\rm min}, v_{\rm max})$ is to clip value $v$ into range $[v_{\rm min}, v_{\rm max}]$. ${\rm round}(v)$ function returns a integer number that is a rounded version of the specified number $v$. ${\rm card}(\boldsymbol{S})$ returns the number of elements in the multiset $\boldsymbol{S}$. $\mathrm{sign}(v)$ returns sign of $v$.

\section{Preliminaries}\label{sec: Preliminaries}
\subsection{MIMO System Model}
We consider a narrow-band MIMO communication system with $N_t$ transmitting (Tx) antennas and $N_r$ receiving (Rx) antennas. Suppose the modulation mode is $Q$-QAM with constellation $\overline{\Omega}$. The system model is as follows,
\begin{equation}\label{Eq:mimo_model}
	\overline{\mathbf{y}} = \overline{\mathbf{H}} \overline{\mathbf{x}} + \overline{\mathbf{n}},
\end{equation}
where $\overline{\mathbf{y}} \in \mathbb{C}^{N_r\times 1}$ and $\overline{\mathbf{x}} \in \overline{\Omega}^{N_t\times 1}$ are the received and transmitted vectors, respectively. $\overline{\mathbf{H}} \in \mathbb{C}^{N_r \times N_t}$ is the channel matrix. The independent and identically distributed (i.i.d.) Rayleigh channel with mean zero and variance $1/N_r$ is assumed in this paper. $\overline{\mathbf{n}}\sim \mathcal{CN}\!\left(0, \overline{\sigma}_n^2\mathbf{I}_{N_r}\right)$ is the additive white \emph{Gaussian} noise (AWGN) vector with mean zero and variance $\overline{\sigma}_n^2$. We assume the channel state information (CSI) is perfectly known at the receiving end.
The complex model is regularly transformed into an equivalent real model following \cite{haykin2004turbo},
\begin{equation}\label{Eq:mimo_model_real}
	\mathbf{y} = \mathbf{H} \mathbf{x} + \mathbf{n},
\end{equation}
where $\mathbf{y} \in \mathbb{R}^{2N_r\times 1}$, $\mathbf{H} \in \mathbb{R}^{2N_r\times 2N_t}$ and $\mathbf{x} \in \Omega^{2N_t\times 1}$. $\Omega$ is the set of real/imaginary part of $Q$-QAM constellation with size of $\sqrt{Q}$. $\mathbf{n}\sim \mathcal{N}\!\left(0, \sigma_n^2\mathbf{I}_{2N_r}\right)$ in which $\sigma_n^2=\overline{\sigma}_n^2/2$.

\subsection{Deep Reinforcement Learning}
DRL, as a branch of machine learning (ML), is concerned with an agent interacting with the environment to make good decisions, in which the Markov decision process (MDP) is assumed \cite{graesser2019foundations}. At timestep $t$, the agent receives a state $s_t\in \mathcal{S}$ by interacting with the environment and then selects an action $a_t\in \mathcal{A}$ according to policy $\pi:\mathcal{S}\to\mathcal{A}$. The environment returns a reward $r_t=\mathcal{R}(s_t,a_t,s_{t+1})$ and the next state $s_{t+1}$ after accepting the agent's action. The above process repeats until the environment returns termination, which is called an episode. Here, $\mathcal{S}$ and $\mathcal{A}$ indicate state space and action space, respectively. $\mathcal{R}$ is the reward function.

Proximal policy optimization (PPO) has become one of the most popular policy gradient algorithms due to its easy implementation and inexpensive computation. Trust region policy optimization (TRPO) is first proposed to solve the sample-inefficient problem in on-policy DRL algorithms by using importance sampling \cite{schulman2015trust}. Later, PPO is proposed to improve the performance and reduce the complexity of TRPO by modifying the objective function \cite{schulman2017proximal}.
For more detailed theories of PPO, we recommend that readers refer to \cite{schulman2017proximal}.

\subsection{Quantization}
Linear quantization is adopted in this paper due to its efficient implementation on hardware. Specifically, the quantization scheme is as follows, the sign, integral part, and fractional part take $1$, $p$, and $q$, respectively, abbreviated as $1-p-q$. For a variable with the value of $v$, the quantized value $v_{\rm Q}$ can be expressed as:
\begin{equation}\label{Eq:quan}
	v_{\rm Q}={\rm round}({\rm clip}(v, {\rm B_{min}}, {\rm B_{max}})/{\rm C})\times{\rm C},
\end{equation}
where ${\rm B_{min}}=-2^p$, ${\rm B_{max}}=2^p-2^{-q}$ and ${\rm C}=2^{-q}$.

\section{The Proposed AHPQ}\label{sec: DRLQ}
For complexity consideration, we consider IPQ and FPQ to be conducted separately.
Since the integral part bitwidth can be determined by the PDF of the data, we can generate a large amount of data to obtain the statistics of variables.

Denote the data generated by Monte Carlo simulation as a multiset $\boldsymbol{S}$. For $1-p-q$ quantization scheme, the range of variables is $[{\rm B_{min}}, {\rm B_{max}}]$. The multiset composed of the elements in $\boldsymbol{S}$ that is not within the range $[{\rm B_{min}}, {\rm B_{max}}]$ is denoted as $\boldsymbol{S}'$, and $\boldsymbol{S}'=\{v|v<{\rm B_{min}}\text{ or }v>{\rm B_{max}},v\in \boldsymbol{S}\}$. The integral part bitwidth can be determined as follows:
\begin{equation}\label{Eq:Intergal}
	p^*=\min_{p}\left\{\frac{{\rm card}(\boldsymbol{S}')}{{\rm card}(\boldsymbol{S})}\leqslant\varepsilon_1\right\},
\end{equation}
where $\varepsilon_1$ is the threshold. When $\varepsilon_1=0$, ${\rm card}(\boldsymbol{S}')=0$, meaning that the range of the variable can be covered in such quantization scheme.
Since ${\rm B_{max}}$ is related to $q$, we can first determine the fractional part bitwidth given large enough integral part bitwidth. With determined fractional part bitwidth, the final integral part bitwidth can be obtained according to Eq.~(\ref{Eq:Intergal}).

As for FPQ, Monte Carlo simulation is generally used to try out the optimal quantization strategy that meets the performance conditions but with huge time overhead. In the following, we demonstrate the proposed DRL-based FPQ in terms of state, action, and reward function.

\subsection{State}
The state is defined as a vector $(k, q_{k})$, where $k\in\{1,2,...,N_{\rm all}\}$ is the $k$-th variable to be quantized. $N_{\rm all}$ denotes the number of variables to be quantized. $q_{k}\in\{0,1,...,q_{\rm max}\}$ is the fractional part bitwidth of the $k$-th variable and $q_{\rm max}$ denotes the preset maximum value of $q_{k}$. Considering that the two-dimensional state may have poor learning ability for detection algorithms with many variables, we instead use one-hot encoding for $k$ and $q_{k}$. Thus the state becomes $(\textup{oh}[k], \textup{oh}[q_{k}])$, where operation $\textup{oh}[\cdot]$ returns the one-hot encoding value. Then the dimension of state becomes $N_{\rm all}+q_{\rm max}+1$.

\subsection{Action}
The action $a_t$ indicates the amount of change in fractional part bitwidth. Inspired by the circular queue mechanism, we use the following formula of fractional part bitwidth change,
\begin{equation}
	q'_{k}\equiv q_{k}+a_t\mod (q_{\rm max}+1),
\end{equation}
where $q'_{k}$ is the fractional part bitwidth after taking action. The reason we do not clip $q'_{k}$ into range $[0, q_{\rm max}]$ is that clipping operation will increase the probability of the agent getting stuck at $0$ or $q_{\rm max}$ (values less than $0$ or greater than $q_{\rm max}$ are forced to be $0$ or $q_{\rm max}$, respectively).
After the agent takes action $a_t$, the fractional part bitwidth of $k$-th variable is changed to $q'_{k}$.

\begin{figure}[htbp]
	\centering
	\subfloat[$L_a=1$ \label{fig:LoS1}]{\includegraphics[width=0.3\columnwidth]{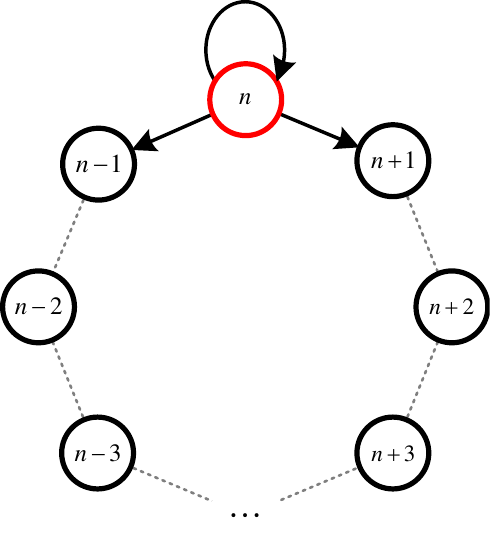}}\ \
	\subfloat[$L_a=2$ \label{fig:LoS2}]{\includegraphics[width=0.3\columnwidth]{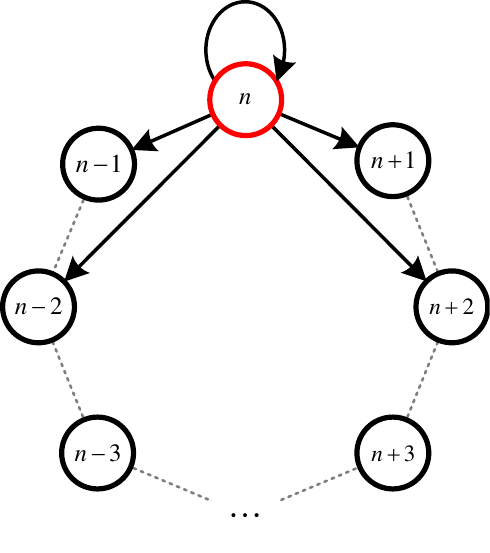}}\ \
	\subfloat[$L_a=\lfloor{q_{\rm max}}/{2}\rfloor$ \label{fig:LoSall}]{\includegraphics[width=0.3\columnwidth]{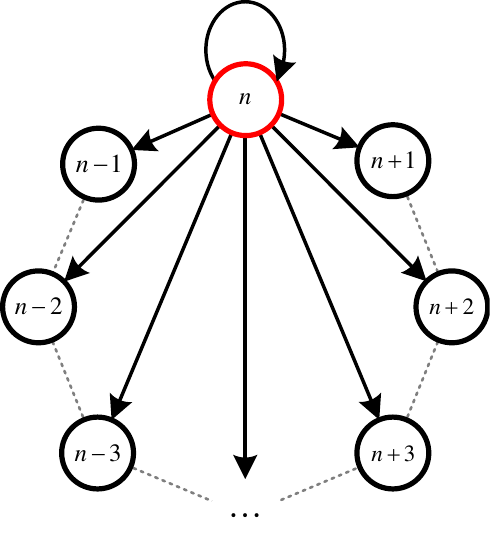}}
	\caption{Schematic diagram of action with different $L_a$.}
	\label{fig:LoS}
\end{figure}
We define the influence range of an action as the maximum absolute value of bitwidth change at an action, referred to as $L_a$. Thus, the action space $\mathcal{A}$ is the set $\{a|-L_a\leqslant a\leqslant L_a, a\in\mathbb{Z}\}$. The schematic diagram of action space with different $L_a$ is presented in Fig.~\ref{fig:LoS}. When $L_a=1$, the bitwidth can only be changed by at most $1$ bit, which may cause low learning efficiency and easy to fall into local optima. When $L_a=\lfloor{q_{\rm max}}/{2}\rfloor$, the agent can flexibly change current bitwidth to any other bitwidth. However, this may lead to instability in learning process and deterioration in convergence performance.

\subsection{Reward Function}
Due to the strict performance requirements in MIMO detection, the reward function in our DRL-based FPQ aims at reducing the fractional part bitwidth under the premise of ensuring the BER performance.

We first focus on the BER performance evaluation. Since the precise BER requires large number of samples for Monte Carlo simulation to approach, it can be pre-computed before the agent starts learning, which is referred to as $P_b$. Every time the environment functions, a small number of samples is used to evaluate the BER performance of the floating-point detector and quantized detector simultaneously to save the computation time. The corresponding BER of the floating-point detector and the quantized detector is written as $P_b^{\rm FL}$ and $P_b^{\rm Q}$, respectively. The relative error of the BER of quantized detector is defined as $(P_b^{\rm Q}-P_b^{\rm FL})/{P_b^{\rm FL}}$. Considering that $P_b^{\rm FL}$ can be equal to $0$ due to the small number of simulation samples, we instead use $(P_b^{\rm Q}-P_b^{\rm FL})/P_b$ as the relative error. If the relative error is greater than threshold $\varepsilon_2$, the performance of the quantized detector cannot be guaranteed and the environment returns the reward $r_t=-1$. Otherwise, the quantization bitwidth is taken into consideration.

Denote the average fractional part bitwidth of the current quantization scheme as $\bar{q}$. The reward function when relative error is smaller than $\varepsilon_2$ is defined as $r_t=\theta_1\exp(-\theta_2{\bar{q}}/{q_{\rm max}})$, where $\theta_1$ and $\theta_2$ are the empirical parameters. Compared with a linear function, the exponential function can make the environment return a larger reward value when bitwidth is small, causing the agent to be more inclined to reduce the quantization bitwidth. To sum up, the reward function is as follows,
\begin{equation}
	r_t=
	\begin{cases}
		\theta_1\exp(-\theta_2{\bar{q}}/{q_{\rm max}}), &(P_b^{\rm Q}-P_b^{\rm FL})/P_b\leqslant\varepsilon_2,\\
		-1, &{\rm otherwise}.
	\end{cases}
\end{equation}


\subsection{Reward Misjudgment Problem}
\begin{figure}[htbp]
	\centering
	\includegraphics[width=1\columnwidth]{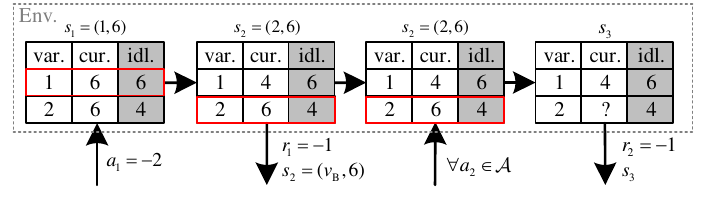}
	\caption{Schematic diagram of a specific case.}
	\label{fig:case}
\end{figure}
Influenced by correlation between different variables in MIMO detection, considering all variables in one episode will introduce severe reward misjudgment and cause poor quantization results. Consider a specific case shown in Fig.~\ref{fig:case}, where Env., var., cur., and idl. denote environment, variable number, current fractional part bitwidth, and ideal fractional part bitwidth, respectively. Suppose the ideal quantization of the two variables $var.\ 1$ and $var.\ 2$ is $var.\ 1:1-1-6$ and $var.\ 2:1-1-4$, and the current quantization of the two variables is $var.\ 1:1-1-6$ and $var.\ 2:1-1-6$. The agent takes an action $a_t=-2$ according to the assumed current state $(1, 6)$, and the quantization of $var.\ 1$ becomes $1-1-4$.
The environment returns $r_t=-1$ and selects $(2,6)$ as the next state. In this case, no matter what action the agent takes, the environment must return $-1$ since the quantization of $var.\ 1$ does not meet the requirements. The reward misjudgment will become more serious for more variables. Therefore, two targeted approaches are proposed to solve the problem.

\subsubsection{Extract Variables}
We extract $N_{\rm ext} (N_{\rm ext}\leqslant N_{\rm all})$ variables randomly that can be adjusted by the agent and other variables are fixed to maximum fractional part bitwidth in each episode. Then the average fractional part bitwidth is defined as follows,
\begin{equation}
	\bar{q}=\frac{1}{N_{\rm ext}}\sum_{k\in \boldsymbol{S}_{\rm ext}}q_{k},
\end{equation}
where $\boldsymbol{S}_{\rm ext}$ denotes the set of extracted variables. Operating $N_{\rm ext}$ extracted variables in one episode can alleviate the reward misjudgment problem, and the correlation between all variables can be considered as long as the number of episodes is sufficient. When $N_{\rm ext}=1$, the correlation between variables is not considered, thus the BER performance is usually unsatisfactory.

\subsubsection{Random Selection}
After the environment accepting the action, the next variable is chosen randomly from $\boldsymbol{S}_{\rm ext}$. Introducing more randomness to the learning process can reduce the probability of reward misjudgment problem.

\subsection{Overall FPQ Process}
In one episode, first, $N_{\rm ext}$ variables are randomly extracted. Then the agent interacts with environment until reaching the maximum timestep. The state, action and reward are collected and stored in a memory. After several episodes, the agent modifies the policy network and value network based on the experience from the memory using PPO algorithm. Once reaching the maximum episode, the agent stops learning.

Since the policy network cannot reflect the statistical characteristics of the fractional part bitwidth, Monte Carlo simulation is needed to obtain them. Fixing the $k$-th variable, the agent after learning keeps taking actions to change the fractional part bitwidth until reaching the maximum testing time. The mean of bitwidth is calculated as estimation of the expectation of $q_k$.

To sum up, the detailed procedure of the proposed AHPQ is listed in Alg.~\ref{alg:DRLQ}, where $T_{\rm max\_episode}$, $T_{\rm max\_timestep}$, $T_{\rm test}$, and $T_{\rm update}$ denote the number of maximum episode, the number of maximum timestep, the number of times to test the fractional part bitwidth, and update interval, respectively.
\begin{algorithm}[htbp]\small
	\SetKwInOut{Input}{Input}\SetKwInOut{Output}{Output}
	\caption{AHPQ algorithm}
	\label{alg:DRLQ}
	{
		$\texttt{set a large enough }p_{k}\texttt{ for all }k$\;
		\tcp{$\texttt{FPQ: Training phase}$}
		\For{\upshape $t_1=1\;\textbf{to}\;T_{\rm max\_episode}$}
		{
			$\texttt{reset environment}$\;
			$\texttt{extract }N_{\rm ext}\texttt{ variables}$\;
			\For{\upshape $t_2=1\;\textbf{to}\;T_{\rm max\_timestep}$}
			{
				$\texttt{agent interacts with environment}$\;
				$\texttt{collect state, action, and reward}$\;
			}
			\If{\upshape $t_1 \% T_{\rm update}==0$}
			{
				$\texttt{update policy using PPO}$\;
			}
		}
		\tcp{$\texttt{FPQ: Testing phase}$}
		\For{\upshape $k=1\;\textbf{to}\;N_{\rm all}$}
		{
			$\texttt{reset environment}$\;
			$\texttt{fix the }k\texttt{-th variable}$\;
			\For{\upshape $t_3=1\;\textbf{to}\;T_{\rm test}$}
			{
				$\texttt{agent takes an action}$\;
				$\texttt{collect state}$\;
			}
			$\texttt{calculate the mean of bitwidth as }q_{k}$\;
		}
		\tcp{$\texttt{IPQ}$}
		$\texttt{obtain }p_{k}\texttt{ according to Eq.~(\ref{Eq:Intergal})}$\;
	}
\end{algorithm}

\begin{Rem}
	The detailed comparison between our proposed AHPQ and other DRL-based CNN quantization, including HAQ \cite{wang2019haq}, ReLeQ \cite{elthakeb2020releq}, and AutoQ \cite{lou2020autoq} are listed in Table~\ref{tab:quan comparison}.
	\begin{table}[htbp]
		\renewcommand{\arraystretch}{1.2}
		\centering
		\caption{The comparison of AHPQ and other DRL-based quantization.}
		\begin{tabular}{p{4.0cm}||p{0.7cm}<{\centering}|p{0.7cm}<{\centering}|p{0.7cm}<{\centering}|p{0.7cm}<{\centering}}
			\Xhline{1.0pt}
			\makecell[c]{Feature} & AHPQ & HAQ & ReLeQ & AutoQ \\
			\hline
			separated IPQ and FPQ & $\checkmark$ & $\times$ & $\times$ & $\times$ \\
			\hline
			flexible action space & $\checkmark$ & $\times$ & $\times$ & $\times$ \\
			\hline
			hardware overhead estimator & $\times^{\text{\ding{172}}}$ & $\checkmark$ & $\times$ & $\checkmark$ \\
			\hline
			solving reward misjudgment & $\checkmark$ & $\times$ & $\times$ & $\times$ \\
			\Xhline{1.0pt}
		\end{tabular}%
		\begin{tablenotes}
			\item[1] \scriptsize{$^{\text{\ding{172}}}$ Due to the varied structure of detectors, the hardware overhead is hard to evaluate and will be explored in our further works.}
		\end{tablenotes}
		\label{tab:quan comparison}
	\end{table}
\end{Rem}

\section{An Application Example}\label{sec: Application Examples}
\subsection{NNA-AMP Detector}
AMP, one popular BMP algorithm, has been widely considered in MIMO detection for a good trade-off between performance and complexity.
The AMP detector is summarized in Alg.~\ref{alg:AMP},
where $\mathbf{b}=\mathbf{H}^{\mathsf{T}}\mathbf{y}$ denotes the received vector after matched filter, $\mathbf{G}=\mathbf{H}^{\mathsf{T}}\mathbf{H}$ is the Gram matrix, $b_i$ is the $i$-th element of $\mathbf{b}$, and $g_{i,j}$ is the $(i,j)$-th element of $\mathbf{G}$. $E_s$ denotes the mean symbol energy. $[\omega_1,...,\omega_{\sqrt{Q}}]$ is the elements of $\Omega$ in ascending order. $u_{i}^{(l)}=\sum_{m'=1}^{\sqrt{Q}}{\exp\!\left[\Delta_i^{(l)}(\omega_{m'})\right]}$ is the normalization coefficient. We recommend readers to refer to \cite{maleki2013asymptotic, jeon2015optimality} for the detailed derivation and analysis of AMP algorithm.

For high order modulation, AMP detector suffers from unaffordable computational complexity due to the moment matching (Lines $6$-$8$ in Alg.~\ref{alg:AMP}). Nearest-neighbor approximation (NNA) proposed in \cite{tang20210} simplifies the moment matching process by choosing the $N_{\Omega}$ nearest neighbor symbols in constellation to compute the probabilities $\rho_i^{(l)}(\omega_m)$. In particular, if $N_{\Omega}=2$, the probabilities $\rho_i^{(l)}(\omega_m)$ can be simplified as follows,
\begin{equation}\label{Eq:probabilities}
	\rho_i^{(l)}(\omega_m)=\left\{
	\begin{aligned}
		&\frac{1}{1+\exp(\Delta_i^{(l)})}, &m=m_1,\\
		&1-\rho_i^{(l)}(\omega_{m_1}), &m=m_2,\\
		&0, &{\rm otherwise}.
	\end{aligned}
	\right.
\end{equation}
and
\begin{equation}\label{Eq:Delta}
	\Delta_i^{(l)} = -\left|\frac{s_{\omega}\left(2z_i^{(l)}-a_{\omega}\right)}{2\tau^{(l)}}\right| = -\left|s_{\omega}\chi_i^{(l)}-\frac{a_{\omega}s_{\omega}}{2\tau^{(l)}}\right|,
\end{equation}
where $a_{\omega}=\omega_{m_1}+\omega_{m_2}$, $s_{\omega}=\omega_{m_2}-\omega_{m_1}$, $\chi_i^{(l)}=z_i^{(l)}/\tau^{(l)}$, $m_1$ and $m_2$ denote the index of the first and second nearest neighbor symbols in constellation, respectively.
In the following, AMP with $N_{\Omega}=2$ NNA is adopted, referred to as NNA-AMP.

\begin{algorithm}[htbp]\small
	\SetKwInOut{Input}{Input}\SetKwInOut{Output}{Output}
	\caption{AMP Algorithm}
	\label{alg:AMP}
	\Input{$\mathbf{b}, \mathbf{G}, \sigma_n^2, L,$ \\
		${d}_i^{(0)}={b}_i, \hat{x}_i^{(0)}=E_s, \bar{\xi}^{(0)}=0 (\forall i=1,...,2N_t)$.}
	\Output{$\hat{x}_{i}^{(L)}$ $(\forall i=1,...,2N_t)$.}
	
	\For{$l = 0, 1, ... , L-1$}
	{
		\For{$i = 1, 2, ... , 2N_t$}
		{
			$z_i^{(l)}=\hat{x}_i^{(l)}+d_i^{(l)}$;\\
			$\tau^{(l)}=\sigma_n^2+\beta\bar{\xi}^{(l)}$;\\
			\For{$m = 1, 2, ... , \sqrt{Q}$}
			{
				$\alpha_{i}^{(l)}(\omega_m)=-{(\omega_m-z_i^{(l)})^2}/{2\tau^{(l)}}$;\\
				$\Delta_i^{(l)}(\omega_m)=\alpha_{i}^{(l)}(\omega_m)-\max_{\omega_{m'}\in\Omega}{\alpha_{i}^{(l)}(\omega_{m'})}$;\\
				$\rho_i^{(l+1)}(\omega_m)=\frac{1}{u_{i}^{(l)}}\exp\!\left[\Delta_i^{(l)}(\omega_m)\right]$;\\
			}
			$\hat{x}_i^{(l+1)}=\sum_{m'=1}^{\sqrt{Q}}\omega_{m'}\rho_i^{(l+1)}(\omega_{m'})$;\\
			$\xi_i^{(l+1)}=\sum_{m'=1}^{\sqrt{Q}}\omega_{m'}^2\rho_i^{(l+1)}(\omega_{m'})-(\hat{x}_i^{(l+1)})^2$;\\
		}
		$\bar{\xi}^{(l+1)}=\sum_{i'=1}^{2N_t}\xi_{i'}^{(l+1)}$;\\
		\For{$i = 1, 2, ... , 2N_t$}
		{
			$d_i^{(l+1)}=b_i-\sum_{j'=1}^{2N_t}g_{i,j'}\hat{x}_{j'}^{(l+1)}+\frac{\beta\bar{\xi}^{(l+1)}}{\tau^{(l)}}d_i^{(l)}$;
		}
	}
\end{algorithm}

\subsection{Implementation Details} \label{subsec: Implementation Details}
\begin{figure}[htbp]
	\centering
	\usepgfplotslibrary{groupplots}
\begin{tikzpicture}
\definecolor{myorange}{RGB}{255,165,0}
\definecolor{myred}{RGB}{247,25,25}
\definecolor{mypink}{RGB}{0,191,255}
\definecolor{myyellow}{RGB}{237,137,32}
\definecolor{mypurple}{RGB}{126,47,142}
\definecolor{myblues}{RGB}{77,190,238}
\definecolor{mygreen}{RGB}{32,134,48}
  \pgfplotsset{
    label style = {font=\fontsize{9pt}{7.2}\selectfont},
    tick label style = {font=\fontsize{7pt}{7.2}\selectfont}
  }

\usetikzlibrary{
    matrix,
}
\begin{axis}[
	scale = 1,
    ymode=log,
    xlabel={Iterations}, xlabel style={yshift=0.6em},
    ylabel={BER}, ylabel style={yshift=-0.65em},
    grid=both,
    ymajorgrids=true,
    xmajorgrids=true,
    xmin=1,xmax=10,
    ymin=1E-04,ymax=1E-01,
    ytick={1E-04,2E-04,3E-4,4E-4,5E-4,6E-4,7E-4,8E-4,9E-4,1E-03,2E-03,3E-3,4E-3,5E-3,6E-3,7E-3,8E-3,9E-3,1E-02,2E-02,3E-2,4E-2,5E-2,6E-2,7E-2,8E-2,9E-2,1E-01},
    yticklabels={$\text{10}^{\text{-4}}$,$$,$$,$$,$$,$$,$$,$$,$$,$\text{10}^{\text{-3}}$,$$,$$,$$,$$,$$,$$,$$,$$,$\text{10}^{\text{-2}}$,$$,$$,$$,$$,$$,$$,$$,$$,$\text{10}^{\text{-1}}$},
    grid style=dashed,
    xshift=-1.8\columnwidth,
    width=1\columnwidth, height=0.6\columnwidth,
    thick,
    legend style={
      nodes={scale=1, transform shape},
      legend columns=1,
      at={(1.12,-0.25)},
      anchor={center},
      cells={anchor=west},
      column sep= 2.5mm,
      row sep= -0.25mm,
      font=\fontsize{7.5pt}{7.2}\selectfont,
    },
	legend pos=north east,
    ]


\addplot[
    color=myred,
    mark=*,
    fill opacity=0,
    line width=0.5mm,
    mark size=2,
]
table {
1	0.0495328125000000
2	0.000923437500000000
3	0.000279687500000000
4	0.000249062500000000
5	0.000235312500000000
6	0.000228437500000000
7	0.000222187500000000
8	0.000221562500000000
9	0.000220312500000000
10	0.000222187500000000
};
\addlegendentry{AMP}

\addplot[
	color=mygreen,
	mark=diamond*,
	densely dashed,
	every mark/.append style={solid},
	fill opacity=0,
	line width=0.5mm,
	mark size=2.5,
]
table {
1	0.0495328125000000
2	0.00128312500000000
3	0.000294687500000000
4	0.000259375000000000
5	0.000245937500000000
6	0.000240000000000000
7	0.000236562500000000
8	0.000234687500000000
9	0.000232812500000000
10	0.000230937500000000
};
\addlegendentry{NNA-AMP}
\end{axis}

\end{tikzpicture}
	\caption{BER performance of different algorithms versus iterations at $6$ dB.}
	\label{fig:BERvsIter}
\end{figure}
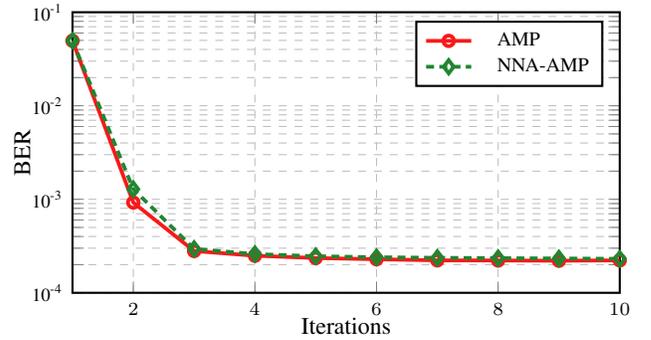

We consider a massive MIMO system with $N_t=8$ and $N_r=128$. The modulation mode is $16$-QAM. In NNA-AMP algorithm, all variables and their corresponding number are listed in Table \ref{tab:variable}.
\begin{table}[htbp]
	\tabcolsep 1mm
	\renewcommand{\arraystretch}{1.2}
	\centering
	\footnotesize
	\caption{Variables and their corresponding numbers.}
	\label{tab:variable}
	\begin{tabular}{p{0.6cm}<{\centering}|p{3.2cm}<{\centering}|p{0.6cm}<{\centering}|p{3.2cm}<{\centering}}
		\Xhline{1.0pt}
		{$\boldsymbol{k}$} & {$\textbf{Variable}^{\text{a}}$} & {$\boldsymbol{k}$} & {$\textbf{Variable}^{\text{a}}$}  \\
		\Xhline{1.0pt}\Xhline{1.0pt}
		$1$ & $b_i$ & $12$ & $\beta\bar{\xi}^{(l)}$ \\ \hline
		$2$ & $g_{i,j}$ & $13$ & $\tau^{(l)}$ \\ \hline
		$3$ & $\sigma^2_n$ & $14$ & ${1}/{\tau^{(l)}}$ \\ \hline
		$4$ & $\rho^{(l)}_{i}(\omega_m)$ & $15$ & $d_i^{(l)}$ \\ \hline
		$5$ & $\omega_m\rho^{(l)}_{i}(\omega_m)$ & $16$ & $z_i^{(l)}$ \\ \hline
		$6$ & $\hat{x}_{i}^{(l)}$ & $17$ & $g_{i,j}\hat{x}_j^{(l)}$ \\ \hline
		$7$ & $(\hat{x}_{i}^{(l)})^2$ & $18$ & $\sum_{j'=1}^{2N_t}g_{i,j'}\hat{x}_{j'}^{(l)}$ \\ \hline
		$8$ & $\omega_m^2\rho^{(l)}_{i}(\omega_m)$ & $19$ & ${\beta\bar{\xi}^{(l)}}/{\tau^{(l)}}$ \\ \hline
		$9$ & $\sum_{m'=1}^{\sqrt{Q}}\omega_{m'}^2\rho^{(l)}_{i}(\omega_{m'})$ & $20$ & $\chi_i^{(l)}$ \\ \hline
		$10$ & $\xi_i^{(l)}$ & $21$ & $\Delta_{i}^{(l)}$ \\ \hline
		$11$ & $\bar{\xi}^{(l)}$ & $-$ & $-$ \\ \hline
		\Xhline{1.0pt}
	\end{tabular}
	\begin{tablenotes}
		\item[1] \hspace{1mm} \scriptsize{$^{\text{a}}$ $\forall i,j\in\{1,...,2N_t\}, \forall m\in\{1,...,\sqrt{Q}\}, \forall l\in\{1,...,L\}$.}
	\end{tablenotes}
\end{table}
The maximum iteration number $L$ is set to $4$ according to Fig.~\ref{fig:BERvsIter}.

Fixed throughout the experiment, $\varepsilon_1=10^{-4}$, $\varepsilon_2=0.5$, $\theta_1=10$, and $\theta_2=4$ can provide good trade-off between BER performance and bitwidth after a certain simulation.

Both policy and value networks consist of fully connected DNN with $6$ hidden layers. The dimensions of the $6$ hidden layers are $64$, $128$, $256$, $256$, $128$, $64$, respectively. The dimensions of the input layers of both policy network and value network are equal to the dimension of state space, which is $32$ in our case. The dimension of the output layer of policy network is equal to the dimension of action space, and that of the value network is $1$.

\begin{figure*}[htbp]
	\centering
	\subfloat[\label{fig:Rew-Iter-LOS}]{\includegraphics[width=1\columnwidth]{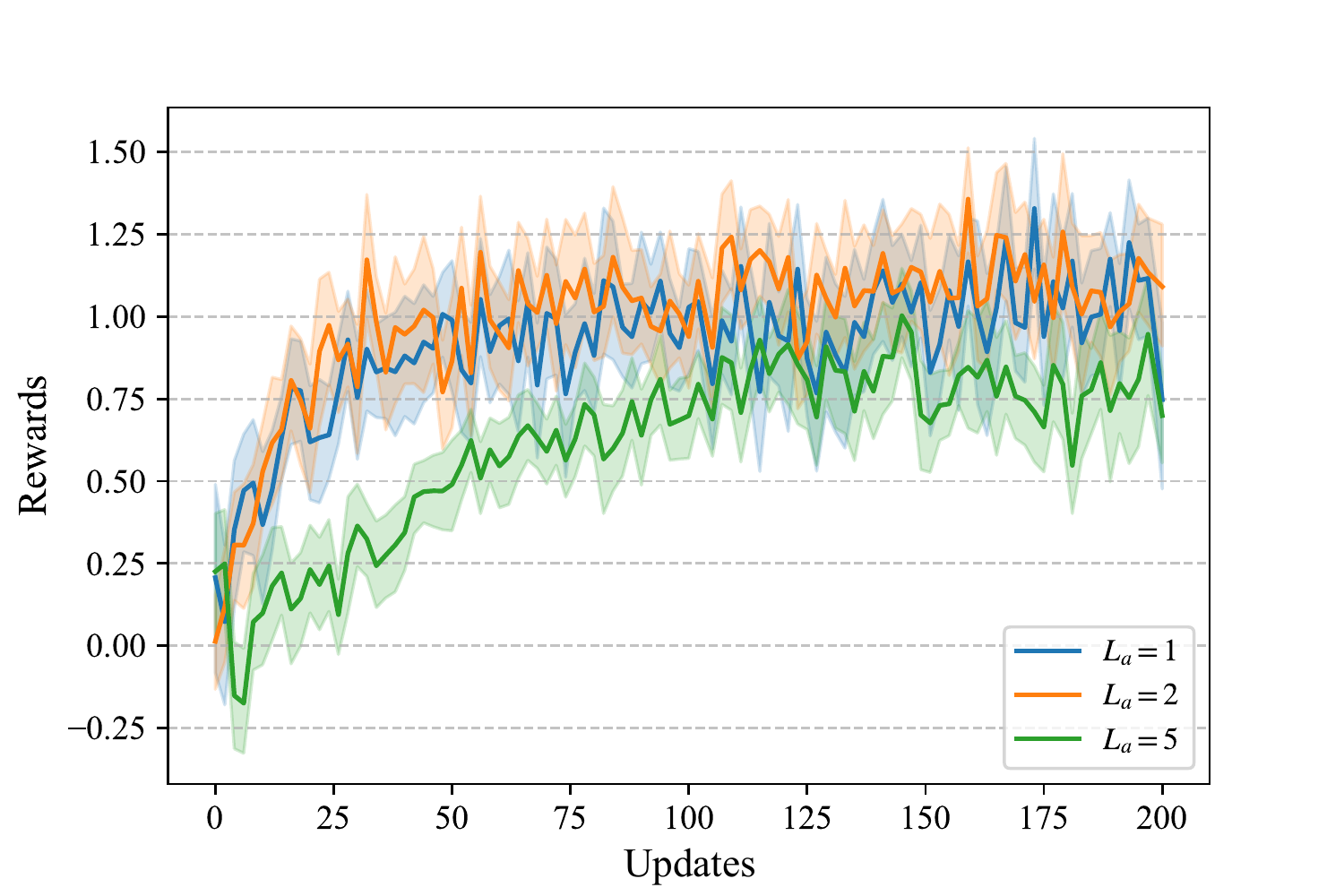}}
	\subfloat[\label{fig:Rew-Iter}]{\includegraphics[width=1\columnwidth]{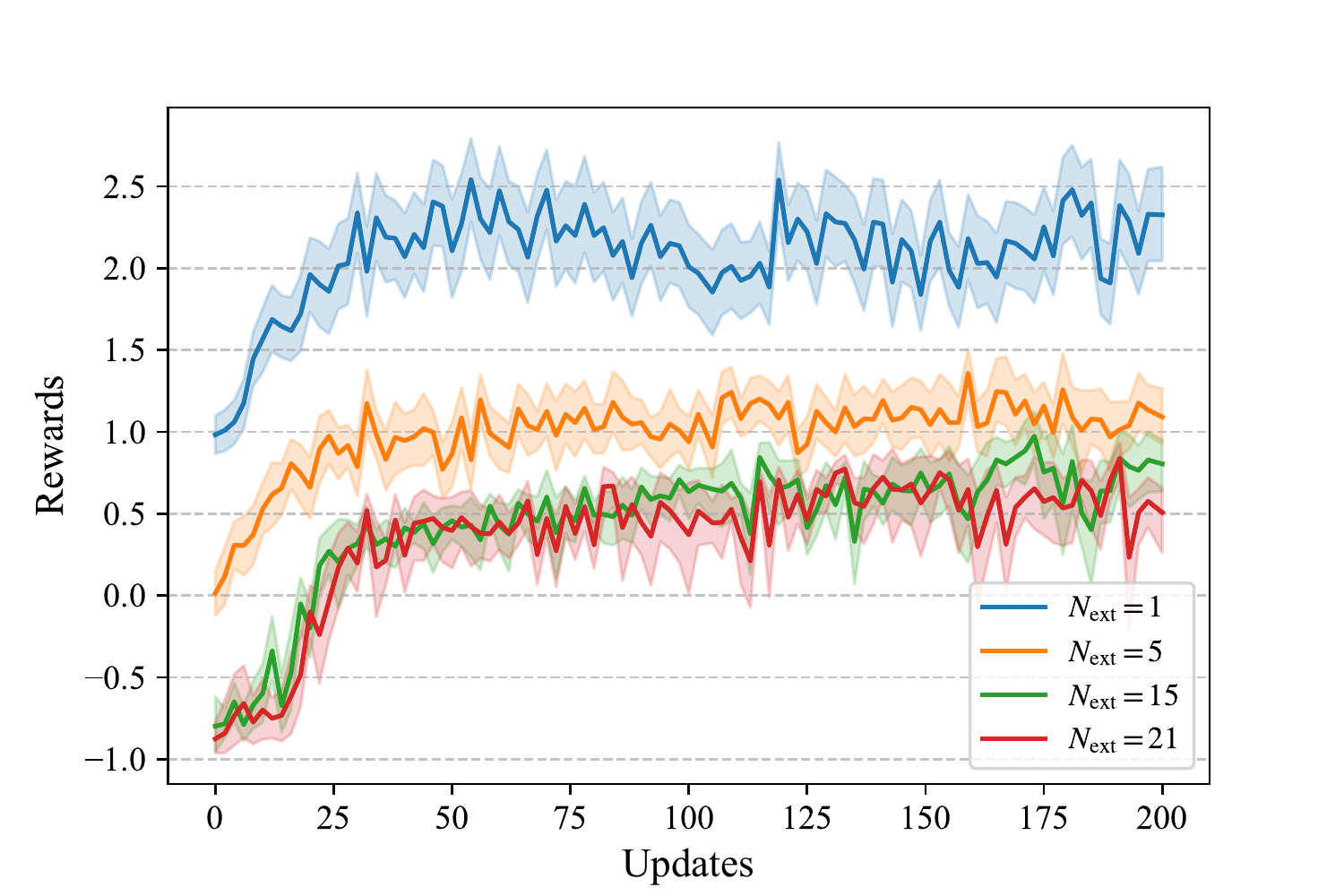}}
	\caption{The rewards versus updates of policy with different (a) $L_a$ and (b) $N_{\rm ext}$.}
\end{figure*}

\subsection{Simulation Results} \label{subsec: Simulation Results}
\subsubsection{Influence of $L_a$}
Fixing $N_{\rm ext}=5$, the rewards versus updates of policy with different $L_a$ is shown in Fig.~\ref{fig:Rew-Iter-LOS}. As illustrated, the convergence performance has a slight advantage when $L_a=2$ compared with the case of $L_a=1$. The reward value eventually converges to approximately the same value in both cases $L_a=1$ and $L_a=2$. When $L_a=5$, the convergence rate becomes very slow, and the reward value at convergence is smaller than the other two cases. Therefore, the convergence rate and final fractional part bitwidth are unsatisfactory when $L_a$ is too large. In the following discussion, we fix $L_a$ to $2$.

\subsubsection{Influence of $N_{\rm ext}$}
The rewards returned by environment versus updates of policy with different $N_{\rm ext}$ is shown in Fig.~\ref{fig:Rew-Iter}. As the number of updates increases, the reward value increases first and finally fluctuates around a certain value, indicating the agent has been learning to quantize variables in NNA-AMP detector better. Since the correlation between different variables is not taken into consideration, the reward value is large when $N_{\rm ext}=1$, indicating that the fractional part bitwidth of variables is reduced to a very low level. As $N_{\rm ext}$ increases, the reward value at convergence becomes smaller. This is mainly because the more severe reward misjudgment problem makes the agent more challenging to reduce the fractional part bitwidth.



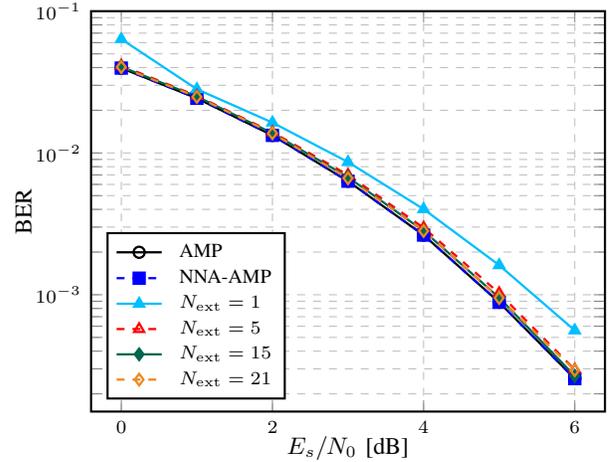
\begin{figure}[htbp]
	\centering
	\usepgfplotslibrary{groupplots}
\begin{tikzpicture}
\definecolor{myblued}{RGB}{0,114,189}
\definecolor{myred}{RGB}{247,25,25}
\definecolor{mypink}{RGB}{0,191,255}
\definecolor{myyellow}{RGB}{237,137,32}
\definecolor{mypurple}{RGB}{200,0,200}
\definecolor{myblues}{RGB}{77,190,238}
\definecolor{mygreen}{RGB}{0,102,80}
  \pgfplotsset{
    label style = {font=\fontsize{9pt}{7.2}\selectfont},
    tick label style = {font=\fontsize{7pt}{7.2}\selectfont}
  }

\usetikzlibrary{
    matrix,
}

\begin{axis}[
	scale = 1,
    ymode=log,
    xlabel={$E_s/N_0$ [\text{dB}]}, xlabel style={yshift=0.6em},
    ylabel={BER}, ylabel style={yshift=-0.55em},
    grid=both,
    ymajorgrids=true,
    xmajorgrids=true,
    xmin=-0.4,xmax=6.4,
    ymin=1.5E-04,ymax=1E-01,
    ytick={2E-04,3E-4,4E-4,5E-4,6E-4,7E-4,8E-4,9E-4,1E-03,2E-03,3E-3,4E-3,5E-3,6E-3,7E-3,8E-3,9E-3,1E-02,2E-02,3E-2,4E-2,5E-2,6E-2,7E-2,8E-2,9E-2,1E-01},
    yticklabels={$$,$$,$$,$$,$$,$$,$$,$$,${10}^{{-3}}$,$$,$$,$$,$$,$$,$$,$$,$$,${10}^{{-2}}$,$$,$$,$$,$$,$$,$$,$$,$$,${10}^{{-1}}$},
    xtick={0, 2, 4, 6},
    xticklabels={${0}$,${2}$,${4}$,${6}$},
    grid style=dashed,
    width=0.95\columnwidth, height=0.78\columnwidth,
    thick,
    legend style={
    	nodes={scale=1, transform shape},
    	legend columns=1,
    	cells={anchor=west},
    	column sep= 1.5mm,
    	row sep= -0.25mm,
    	font=\fontsize{7.5pt}{7.2}\selectfont,
    },
    legend pos=south west,
    ]

\addplot[
    color=black,
    mark=*,
    fill opacity=0,
    line width=0.3mm,
    mark size=2.1,
]
table {
0	0.0396737500000000
1	0.0243000000000000
2	0.0133109375000000
3	0.00632421875000000
4	0.00263515625000000
5	0.000884687500000000
6	0.000256093750000000
};
\addlegendentry{AMP}

\addplot[
    color=blue,
    mark=square*,
    dashed,
    every mark/.append style={solid},
    line width=0.3mm,
    mark size=2.1,
]
table {
0	0.0396745312500000
1	0.0243025000000000
2	0.0133104687500000
3	0.00632437500000000
4	0.00263515625000000
5	0.000884687500000000
6	0.000256093750000000
};
\addlegendentry{NNA-AMP}

\addplot[
    color=mypink,
    mark=triangle*,
    line width=0.3mm,
    mark size=2.2,
]
table {
0	0.0635437500000000
1	0.0281612500000000
2	0.0163775000000000
3	0.00862312500000000
4	0.00401625000000000
5	0.00161609375000000
6	0.000560312500000000
};
\addlegendentry{$N_{\rm ext}=1$}

\addplot[
    color=red,
    mark=triangle*,
    fill opacity=0,
    dashed,
    every mark/.append style={solid},
    line width=0.3mm,
    mark size=2.2,
]
table {
0	0.0412926562500000
1	0.0252512500000000
2	0.0139946875000000
3	0.00700250000000000
4	0.00296687500000000
5	0.00102578125000000
6	0.000296093750000000
};
\addlegendentry{$N_{\rm ext}=5$}

\addplot[
    color=mygreen,
    mark=diamond*,
    line width=0.3mm,
    mark size=2.2,
]
table {
0	0.0404795312500000
1	0.0249932812500000
2	0.0137834375000000
3	0.00674234375000000
4	0.00282250000000000
5	0.000952343750000000
6	0.000267656250000000
};
\addlegendentry{$N_{\rm ext}=15$}

\addplot[
    color=myyellow,
    mark=diamond*,
    fill opacity=0,
    dashed,
    every mark/.append style={solid},
    line width=0.3mm,
    mark size=2.2,
]
table {
0	0.0403396875000000
1	0.0248746875000000
2	0.0137703125000000
3	0.00662875000000000
4	0.00281781250000000
5	0.000948750000000000
6	0.000293593750000000
};
\addlegendentry{$N_{\rm ext}=21$}
\end{axis}

\end{tikzpicture}
	\caption{BER performance comparison of AMP, NNA-AMP, and AHPQ-based quantized NNA-AMP with different $N_{\rm ext}$.}
	\label{fig:BER-QS}
\end{figure}

The BER performance comparison of AMP, NNA-AMP, and AHPQ-based quantized NNA-AMP with different $N_{\rm ext}$ is shown in Fig.~\ref{fig:BER-QS}, where the legend entry ``$N_{\rm ext}=1$" denotes AHPQ-based quantized NNA-AMP with $N_{\rm ext}=1$ and others are similar. As presented, the NNA-AMP algorithm can perfectly recover the performance of AMP. The quantized NNA-AMP shows degraded performance when $N_{\rm ext}=1$, while it presents the similar performance to the floating-point NNA-AMP when $N_{\rm ext}$ is equal to $5$, $15$, and $21$.

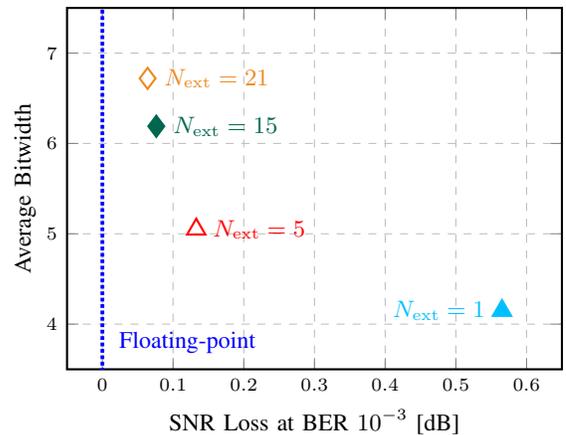
\begin{figure}[htbp]
	\centering
	\usepgfplotslibrary{groupplots}
\usetikzlibrary{arrows.meta}
\begin{tikzpicture}[>=Stealth]
\definecolor{myblued}{RGB}{0,114,189}
\definecolor{myred}{RGB}{247,25,25}
\definecolor{mypink}{RGB}{0,191,255}
\definecolor{myyellow}{RGB}{237,137,32}
\definecolor{mypurple}{RGB}{200,0,200}
\definecolor{myblues}{RGB}{77,190,238}
\definecolor{mygreen}{RGB}{0,102,80}
  \pgfplotsset{
    label style = {font=\fontsize{9pt}{7.2}\selectfont},
    tick label style = {font=\fontsize{7pt}{7.2}\selectfont}
  }

\usetikzlibrary{
    matrix,
}
\begin{axis}[
	scale = 2,
    xlabel={SNR Loss at BER $10^{-3}$ [\text{dB}]}, xlabel style={yshift=0.0em},
    ylabel={Average Bitwidth}, ylabel style={yshift=-0.3em},
    grid=both,
    ymajorgrids=true,
    xmajorgrids=true,
    xmin=-0.05,xmax=0.65,
    ymin=3.5,ymax=7.5,
    ytick={4,5,6,7},
    yticklabels={$4$,$5$,$6$,$7$},
    xtick={0,0.1,0.2,0.3,0.4,0.5,0.6},
    xticklabels={$0$,$0.1$,$0.2$,$0.3$,$0.4$,$0.5$,$0.6$},
    grid style=dashed,
    width=0.55\columnwidth, height=0.45\columnwidth,
    thick,
    ]

\addplot[
    color=blue,
    densely dotted,
    line width=0.5mm,
]
table {
0 0
0 10
};
\node[right, align=left]
at (axis cs:0.01, 3.8) {\fontsize{9}{6}\selectfont{\textcolor{blue}{Floating-point}}};

\addplot[
    color=mypink,
    mark=triangle*,
    mark size=4,
]
table {
0.56525 4.15
};
\node[left, align=left]
at (axis cs:0.55525, 4.15) {\fontsize{9}{6}\selectfont{\textcolor{mypink}{$N_{\rm ext}=1$}}};

\addplot[
    color=red,
    mark=triangle*,
    fill opacity=0,
    mark size=4,
]
table {
0.13275 5.05
};
\node[right, align=left]
at (axis cs:0.14275, 5.05) {\fontsize{9}{6}\selectfont{\textcolor{red}{$N_{\rm ext}=5$}}};

\addplot[
    color=mygreen,
    mark=diamond*,
    mark size=4,
]
table {
0.07625 6.19
};
\node[right, align=left]
at (axis cs:0.08625, 6.19) {\fontsize{9}{6}\selectfont{\textcolor{mygreen}{$N_{\rm ext}=15$}}};

\addplot[
    color=myyellow,
    mark=diamond*,
    fill opacity=0,
    mark size=4,
]
table {
0.06395 6.72
};
\node[right, align=left]
at (axis cs:0.07395, 6.72) {\fontsize{9}{6}\selectfont{\textcolor{myyellow}{$N_{\rm ext}=21$}}};

\end{axis}

\end{tikzpicture}
	\caption{The trade-off between average bitwidth and SNR loss at BER = $10^{-3}$ with the floating-point NNA-AMP as the benchmark.}
	\label{fig:BWvsBER}
\end{figure}

The trade-off between average bitwidth and performance loss after quantization is shown in Fig.~\ref{fig:BWvsBER}. The horizontal axis stands for the SNR loss of NNA-AMP using AHPQ with the floating-point NNA-AMP as the benchmark. The vertical axis represents the average bitwidth. The quantized NNA-AMP when $N_{\rm ext}=5$ can allocate small quantization bitwidth while maintaining the performance of the original algorithm. Thus in the following, $N_{\rm ext}$ is fixed to $5$ and the detailed quantization bitwidth of AHPQ-based quantized NNA-AMP with $N_{\rm ext}=5$ is listed in Table~\ref{tab:quan}.


\begin{figure}[htbp]
	\centering
	\usepgfplotslibrary{groupplots}
\begin{tikzpicture}
\definecolor{myblued}{RGB}{0,114,189}
\definecolor{myred}{RGB}{247,25,25}
\definecolor{mypink}{RGB}{0,191,255}
\definecolor{myyellow}{RGB}{237,137,32}
\definecolor{mypurple}{RGB}{200,0,200}
\definecolor{myblues}{RGB}{77,190,238}
\definecolor{mygreen}{RGB}{0,102,80}
  \pgfplotsset{
    label style = {font=\fontsize{9pt}{7.2}\selectfont},
    tick label style = {font=\fontsize{7pt}{7.2}\selectfont}
  }

\usetikzlibrary{
    matrix,
}

\begin{axis}[
	scale = 1,
    ymode=log,
    xlabel={$E_s/N_0$ [\text{dB}]}, xlabel style={yshift=0.6em},
    ylabel={BER}, ylabel style={yshift=-0.75em},
    grid=both,
    ymajorgrids=true,
    xmajorgrids=true,
    xmin=-0.4,xmax=6.4,
    ymin=1.5E-04,ymax=1E-01,
    ytick={1E-04,2E-04,3E-4,4E-4,5E-4,6E-4,7E-4,8E-4,9E-4,1E-03,2E-03,3E-3,4E-3,5E-3,6E-3,7E-3,8E-3,9E-3,1E-02,2E-02,3E-2,4E-2,5E-2,6E-2,7E-2,8E-2,9E-2,1E-01},
    yticklabels={${10}^{{-4}}$,$$,$$,$$,$$,$$,$$,$$,$$,${10}^{{-3}}$,$$,$$,$$,$$,$$,$$,$$,$$,${10}^{{-2}}$,$$,$$,$$,$$,$$,$$,$$,$$,${10}^{{-1}}$},
    xtick={0, 2, 4, 6},
    xticklabels={${0}$,${2}$,${4}$,${6}$},
    grid style=dashed,
    width=0.95\columnwidth, height=0.78\columnwidth,
    thick,
    legend style={
    	nodes={scale=1, transform shape},
    	legend columns=1,
    	cells={anchor=west},
    	column sep= 1.5mm,
    	row sep= -0.25mm,
    	font=\fontsize{7.5pt}{7.2}\selectfont,
    },
    legend pos=south west,
    ]

\addplot[
    color=black,
    mark=*,
    fill opacity=0,
    line width=0.3mm,
    mark size=2.1,
]
table {
0	0.0398060937500000
1	0.0242734375000000
2	0.0131871875000000
3	0.00634156250000000
4	0.00262656250000000
5	0.000882343750000000
6	0.000239218750000000
};
\addlegendentry{Floating-point}

\addplot[
    color=blue,
    mark=square*,
    line width=0.3mm,
    mark size=2.1,
]
table {
0	0.0466143750000000
1	0.0306721875000000
2	0.0190260937500000
3	0.0106695312500000
4	0.00565937500000000
5	0.00267906250000000
6	0.00123406250000000
};
\addlegendentry{UQ: 1-6-4}

\addplot[
    color=mypink,
    mark=triangle*,
    line width=0.3mm,
    mark size=2.2,
]
table {
0	0.0415707812500000
1	0.0260178125000000
2	0.0147412500000000
3	0.00752750000000000
4	0.00333093750000000
5	0.00129203125000000
6	0.000382031250000000
};
\addlegendentry{UQ: 1-6-5}

\addplot[
    color=red,
    mark=triangle*,
    fill opacity=0,
    dashed,
    every mark/.append style={solid},
    line width=0.3mm,
    mark size=2.2,
]
table {
0	0.0403110937500000
1	0.0247245312500000
2	0.0136242187500000
3	0.00669609375000000
4	0.00282593750000000
5	0.000989843750000000
6	0.000280312500000000
};
\addlegendentry{UQ: 1-6-6}

\addplot[
    color=mygreen,
    mark=diamond*,
    line width=0.3mm,
    mark size=2.2,
]
table {
0	0.0399206250000000
1	0.0243809375000000
2	0.0132614062500000
3	0.00643578125000000
4	0.00267359375000000
5	0.000922812500000000
6	0.000257500000000000
};
\addlegendentry{UQ: 1-6-7}

\addplot[
    color=myyellow,
    mark=diamond*,
    fill opacity=0,
    dashed,
    every mark/.append style={solid},
    line width=0.3mm,
    mark size=2.2,
]
table {
0	0.0398360937500000
1	0.0243084375000000
2	0.0132371875000000
3	0.00636015625000000
4	0.00264437500000000
5	0.000895625000000000
6	0.000246875000000000
};
\addlegendentry{UQ: 1-6-8}
\end{axis}

\end{tikzpicture}
	\caption{BER performance comparison of NNA-AMP and UQ-based quantized NNA-AMP with different quantization.}
	\label{fig:BER-Unity}
\end{figure}
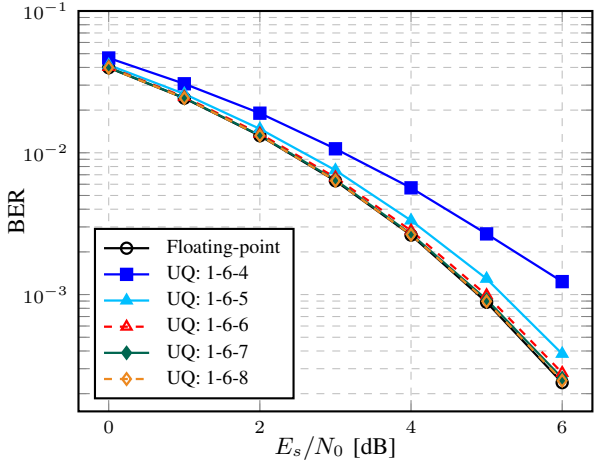

\subsubsection{Comparison with UQ Scheme}
\begin{table*}[htbp]
	\tabcolsep 1mm
	\renewcommand{\arraystretch}{1.2}
	\centering
	\footnotesize
	\caption{Quantization bitwidth comparison of NNA-AMP using AHPQ and UQ.}
	\label{tab:quan}
	\begin{tabular}{cc||p{0.4cm}<{\centering}|p{0.4cm}<{\centering}|p{0.4cm}<{\centering}|p{0.4cm}<{\centering}|p{0.4cm}<{\centering}|p{0.4cm}<{\centering}|p{0.4cm}<{\centering}|p{0.4cm}<{\centering}|p{0.4cm}<{\centering}|p{0.4cm}<{\centering}|p{0.4cm}<{\centering}|p{0.4cm}<{\centering}|p{0.4cm}<{\centering}|p{0.4cm}<{\centering}|p{0.4cm}<{\centering}|p{0.4cm}<{\centering}|p{0.4cm}<{\centering}|p{0.4cm}<{\centering}|p{0.4cm}<{\centering}|p{0.4cm}<{\centering}|p{0.4cm}<{\centering}||p{0.8cm}<{\centering}}
		\Xhline{1.0pt}
		\multicolumn{2}{c||}{$k$} & 1 & 2 & 3 & 4 & 5 & 6 & 7 & 8 & 9 & 10 & 11 & 12 & 13 & 14 & 15 & 16 & 17 & 18 & 19 & 20 & 21 & Avg. \\ \hline
		\multicolumn{1}{c|}{\multirow{2}{*}{Integral Part}}    & AHPQ & \multicolumn{1}{c|}{3} & \multicolumn{1}{c|}{2} & \multicolumn{1}{c|}{1} & \multicolumn{1}{c|}{1} & \multicolumn{1}{c|}{2} & \multicolumn{1}{c|}{2} & \multicolumn{1}{c|}{4} & \multicolumn{1}{c|}{4} & \multicolumn{1}{c|}{4} & \multicolumn{1}{c|}{1}  & \multicolumn{1}{c|}{1}  & \multicolumn{1}{c|}{1}  & \multicolumn{1}{c|}{1}  & \multicolumn{1}{c|}{4}  & \multicolumn{1}{c|}{3}  & \multicolumn{1}{c|}{4}  & \multicolumn{1}{c|}{3}  & \multicolumn{1}{c|}{3}  & \multicolumn{1}{c|}{1}  & \multicolumn{1}{c|}{6}  & 6  & 2.57 \\ \cline{2-24}
		\multicolumn{1}{c|}{}                                 & UQ   & \multicolumn{21}{c||}{6}                                                                                                                                                                                                                                                                                                                                                                                                                                                                                                           & 6    \\ \hline
		\multicolumn{1}{c|}{\multirow{2}{*}{Fractional Part}} & AHPQ & \multicolumn{1}{c|}{6} & \multicolumn{1}{c|}{7} & \multicolumn{1}{c|}{3} & \multicolumn{1}{c|}{3} & \multicolumn{1}{c|}{3} & \multicolumn{1}{c|}{2} & \multicolumn{1}{c|}{1} & \multicolumn{1}{c|}{0} & \multicolumn{1}{c|}{1} & \multicolumn{1}{c|}{0}  & \multicolumn{1}{c|}{0}  & \multicolumn{1}{c|}{1}  & \multicolumn{1}{c|}{3}  & \multicolumn{1}{c|}{1}  & \multicolumn{1}{c|}{4}  & \multicolumn{1}{c|}{4}  & \multicolumn{1}{c|}{6}  & \multicolumn{1}{c|}{4}  & \multicolumn{1}{c|}{1}  & \multicolumn{1}{c|}{1}  & 1  & 2.48 \\ \cline{2-24}
		\multicolumn{1}{c|}{}                                 & UQ   & \multicolumn{21}{c||}{6}                                                                                                                                                                                                                                                                                                                                                                                                                                                                                                           & 6    \\ \Xhline{1.0pt}
	\end{tabular}
\end{table*}
The integral part bitwidth of UQ cannot be less than the maximum integral part bitwidth of all variables, which is $6$ bits. As for fractional part bitwidth, the BER performance comparison of NNA-AMP and NNA-AMP using UQ with different fractional part bitwidths is presented in Fig.~\ref{fig:BER-Unity}.
The NNA-AMP using UQ suffers from severe performance deterioration when fractional part bitwidth is $4$ or $5$ bits, while it can recover the performance of floating-point NNA-AMP when fractional part bitwidth is $6$ bits or more. Therefore, the UQ scheme of NNA-AMP is taken as $1-6-6$. Compared with UQ, AHPQ reduces the bitwidth by $57.2\%$ and $58.7\%$ in integral part and fractional part, respectively.

\section{Hardware-Friendly NNA-AMP Detector and VLSI Architecture}\label{sec: HF AMP and VLSI}
\subsection{Node Compression}\label{s2c: NC}
As summarized in Table~\ref{tab:variable}, there are a large number of processing nodes in the signal flow graph (SFG) of NNA-AMP detector. Benefiting from AHPQ, the detailed quantization information of each variable makes it possible to compress the processing nodes by removing variables with low quantization bitwidth from the SFG.

Consider the operations related to variance updating in Lines $10-13$ of Alg.~\ref{alg:AMP}. Note that the quantization of $\bar{\xi}^{(l)}$ is $1-1-0$ and $\beta=N_t/N_r=1/16$, thus the value of $\beta\bar{\xi}^{(l)}$ must be less than $0.1$. However, the AHPQ claims that the fractional part bitwidth of $\beta\bar{\xi}^{(l)}$ is $1$, leading to this variable being set to $0$ in the quantized NNA-AMP detector. Thus we can simplify Line $4$ and Line $13$ in Alg.~\ref{alg:AMP} as follows,
\begin{subequations}\label{equ:taud}
	\begin{align}
		\tau^{(l)}&=\sigma_n^2,\\
		d_i^{(l+1)}&=b_i-\sum_jg_{i,j}\hat{x}_j^{(l+1)},
	\end{align}
\end{subequations}
and the calculation of ${\xi}_i^{(l)}$ can be removed.

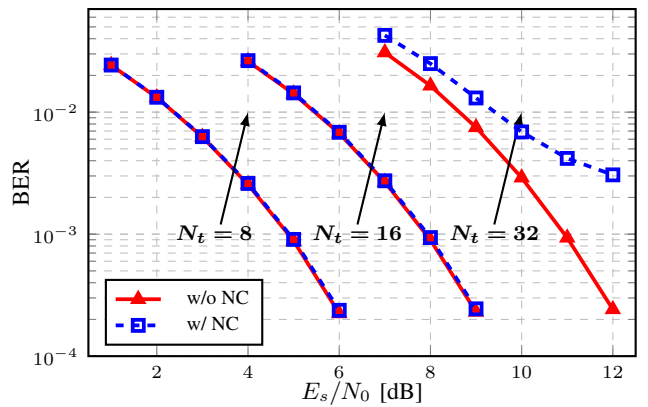
\begin{figure}[htbp]
	\centering
	\usepgfplotslibrary{groupplots}
\begin{tikzpicture}
\definecolor{myblued}{RGB}{0,114,189}
\definecolor{myred}{RGB}{247,25,25}
\definecolor{mypink}{RGB}{0,191,255}
\definecolor{myyellow}{RGB}{237,137,32}
\definecolor{mypurple}{RGB}{200,0,200}
\definecolor{myblues}{RGB}{77,190,238}
\definecolor{mygreen}{RGB}{0,102,80}
  \pgfplotsset{
    label style = {font=\fontsize{9pt}{7.2}\selectfont},
    tick label style = {font=\fontsize{7pt}{7.2}\selectfont}
  }

\usetikzlibrary{
    matrix,
}

\begin{axis}[
	scale = 1,
    ymode=log,
    xlabel={$E_s/N_0$ [\text{dB}]}, xlabel style={yshift=0.6em},
    ylabel={BER}, ylabel style={yshift=-0.55em},
    grid=both,
    ymajorgrids=true,
    xmajorgrids=true,
    xmin=0.5,xmax=12.5,
    ymin=1E-04,ymax=7E-02,
    xtick={0, 2, 4, 6, 8, 10, 12},
    xticklabels={$0$,$2$,$4$,$6$,$8$,$10$,$12$},
    grid style=dashed,
    xshift=-1\columnwidth,
    width=1\columnwidth, height=0.7\columnwidth,
    thick,
    legend style={
      nodes={scale=1, transform shape},
      legend columns=1,
      at={(1.12,-0.25)},
      anchor={center},
      cells={anchor=west},
      column sep= 2.5mm,
      row sep= -0.25mm,
      font=\fontsize{7.5pt}{7.2}\selectfont,
    },
    legend pos=south west,
    ]

\addplot[
	color=red,
	mark=triangle*,
	line width=0.5mm,
	mark size=2.2,
]
table {
	1	0.0242753125000000
	2	0.0131865625000000
	3	0.00625718750000000
	4	0.00255218750000000
	5	0.000880000000000000
	6	0.000224062500000000
};
\addlegendentry{w/o NC}

\addplot[
    color=blue,
    mark=square*,
    dashed,
    every mark/.append style={solid},
    fill opacity=0,
    line width=0.5mm,
    mark size=2.1,
]
table {
	1	0.0243187500000000
	2	0.0132450000000000
	3	0.00629718750000000
	4	0.00260343750000000
	5	0.000905312500000000
	6	0.000237500000000000
};
\addlegendentry{w/ NC}
\node[right, align=left]
at (axis cs:2.2, 1E-03) {\footnotesize{\bm{$N_t=8$}}};
\draw[-latex] (3.4,1.2E-03) -- (4,1E-2);

\addplot[
	color=red,
	mark=triangle*,
	line width=0.5mm,
	mark size=2.2,
]
table {
	4	0.0260864062500000
	5	0.0141420312500000
	6	0.00670546875000000
	7	0.00267406250000000
	8	0.000896250000000000
	9	0.000230468750000000
};

\addplot[
	color=blue,
	mark=square*,
	dashed,
	every mark/.append style={solid},
	fill opacity=0,
	line width=0.5mm,
	mark size=2.1,
]
table {
	4	0.0264357812500000
	5	0.0143837500000000
	6	0.00683640625000000
	7	0.00273828125000000
	8	0.000937656250000000
	9	0.000244375000000000
};
\node[right, align=left]
at (axis cs:5.2, 1E-03) {\footnotesize{\bm{$N_t=16$}}};
\draw[-latex] (6.4,1.2E-03) -- (7,1E-2);

\addplot[
	color=red,
	mark=triangle*,
	line width=0.5mm,
	mark size=2.2,
]
table {
	7	0.0308364843750000
	8	0.0164748437500000
	9	0.00750578125000000
	10	0.00290445312500000
	11	0.000933750000000000
	12	0.000241953125000000
};

\addplot[
	color=blue,
	mark=square*,
	dashed,
	every mark/.append style={solid},
	fill opacity=0,
	line width=0.5mm,
	mark size=2.1,
]
table {
	7	0.0425976562500000
	8	0.0249950781250000
	9	0.0130679687500000
	10	0.00689000000000000
	11	0.00417570312500000
	12	0.00306242187500000
};
\node[right, align=left]
at (axis cs:8.2, 1E-03) {\footnotesize{\bm{$N_t=32$}}};
\draw[-latex] (9.4,1.2E-03) -- (10,1E-2);
\end{axis}

\end{tikzpicture}
	\caption{BER performance comparison of NNA-AMP w/ NC and NNA-AMP w/o NC with $N_r=128$ and different $N_t$.}
	\label{fig:BER-NC}
\end{figure}

To verify the correctness of the aforementioned simplification, we simulate the BER performance of original NNA-AMP and NNA-AMP with NC under different antenna configurations. Fig.~\ref{fig:BER-NC} demonstrates that the system performance only suffers negligible degradation when the antenna ratio $N_t/N_r$ is smaller than $1/8$.

\subsection{Strength Reduction}\label{s2c: SR}
In this subsection, according to the quantization results, we introduce a digital transformation to reorganize the digital signal processing (DSP) operations aiming at reducing the computational burden. In such case, metrics can be efficiently achieved, such as small area, low power, and high throughput.

\subsubsection{Simplification of Mean Calculation}\label{s3c: SS}
We analyze the calculation of $\rho^{(l+1)}_{i}(\omega_m)$ in NNA-AMP with $16$-QAM modulation. How to obtain the positions of the first and second nearest neighbor symbols in constellation via $z_i^{(l)}$ affects the hardware design critically. Searching directly by Euclidean distance requires minus, square, and sorting operations, bringing intolerable latency and complexity.

\begin{figure}[htbp]
	\centering
	\includegraphics[width=1\columnwidth]{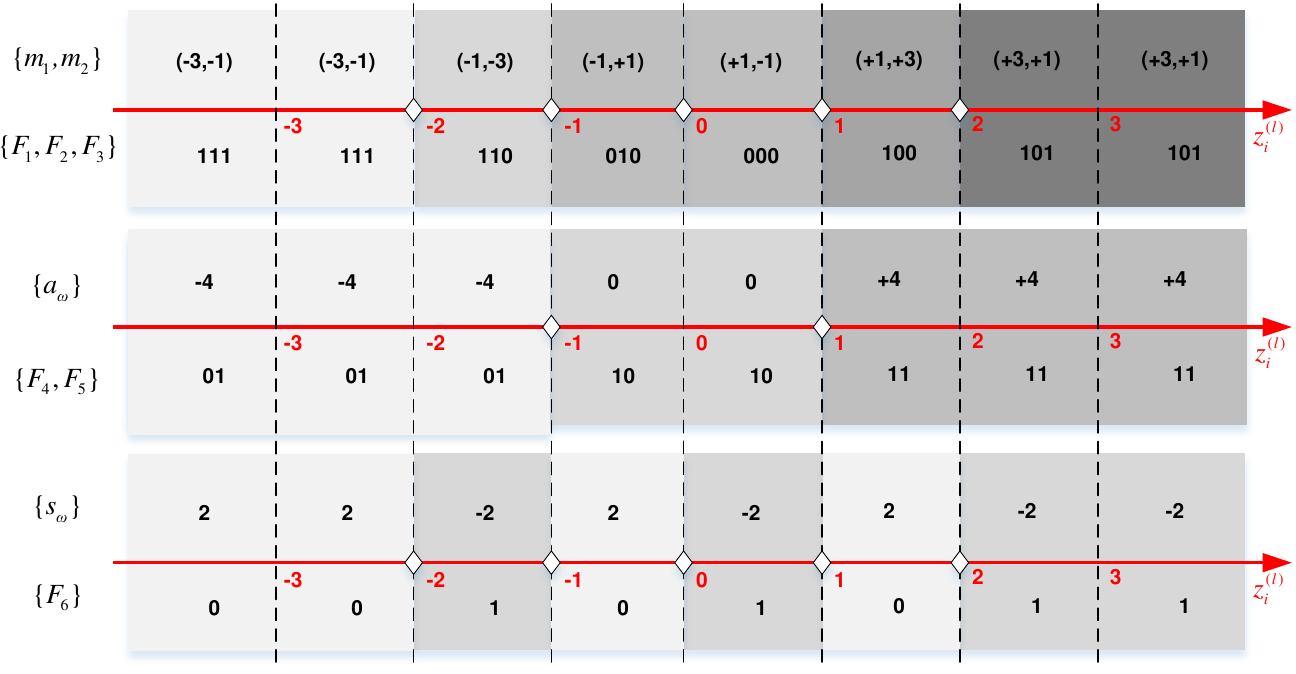}
	\caption{The schematic diagram of obtaining $m_1$, $m_2$, $a_{\omega}$, and $s_{\omega}$ based on the interval division for $16$-QAM NNA-AMP. ($\{F_1,F_2,F_3\}$: flags of choosing $m_1$ and $m_2$; $\{F_4,F_5\}$: flags of $a_{\omega}$ selection; $\{F_6\}$: flag of $s_\omega$ selection. $\Diamond$ represents the boundary of flag selection.)}
	\label{fig:QC}
\end{figure}

Based on the coordinates of constellation, a selective branch structure is chosen to guide the hardware design, which can further simplify the dynamic multiplication associated with $a_{\omega}$ and $s_{\omega}$ in Eq.~(\ref{Eq:Delta}). As indicated in Fig.~\ref{fig:QC}, index $\{m_1, m_2\}$ divides the whole value interval of $z_i^{(l)}$ into six parts as $(-\infty,-2)\cup(-2,-1)\cup(-1,0)\cup(0,1)\cup(1,2)\cup(2,+\infty)$ using Gray-coded flags $\{F_1, F_2, F_3\}$ to distinguish. $a_{\omega}$ and $s_{\omega}$ can be pre-calculated in different intervals. $a_{\omega}$ has three possible values for different intervals, namely $-4, 0, 4$, and the flags $\{F_4, F_5\}$ for choosing different value of $a_{\omega}$ are Gray-coded as $01, 11, 10$. $s_{\omega}$ has two possible values with amplitudes of $-2, 2$, respectively, and the flag $\{F_6\}$ for choosing different value of $s_{\omega}$ is Gray-coded as $0,1$. Hence, Eq.~(\ref{Eq:Delta}) can be simplified as
\begin{equation}\label{Eq:Deltanew}
	\begin{aligned}
		\widetilde{ \Delta}^{(l)}_i&= -2\left|\mathrm{sign}(s_{\omega})\cdot\left(\chi_i^{(l)}-\frac{2}{\tau^{(l)}}\cdot \mathrm{sign}(a_{\omega})\right) \right|\\
		&=-2\left|\chi_i^{(l)}-\frac{2}{\tau^{(l)}}\cdot \mathrm{sign}(a_{\omega})\right|,
	\end{aligned}
\end{equation}
where $-|\cdot|$ and $-\frac{2}{\tau^{(l)}}\cdot \mathrm{sign}(a_{\omega})$ can be obtained by numerical transformation and segment shift selector, respectively. The flag $\{F_6\}$ can be discarded since the calculation of $\widetilde{ \Delta}^{(l)}_i$ is independent of $s_{\omega}$ in Eq.~(\ref{Eq:Deltanew}). Furthermore, considering the calculation of $\hat{x}_i^{(l+1)}$ (Line 9 of Alg.~\ref{alg:AMP}), the multiplication with integerized constellation can be simplified to shift-and-adder (SAA) operations by sub-expression sharing in Table~\ref{tab:mean sel} inspired by \cite{parhi2007vlsi}. With $F_6$ discarded, five flags $\{F_1,F_2,F_3,F_4,F_5\}$ which can be obtained by the input signal $z_i^{(l)}$ directly guide the calculation of $\widetilde{\Delta}^{(l)}_i$ and $\hat{x}_i^{(l+1)}$.

\begin{table}[htbp]
	\renewcommand{\arraystretch}{1.2}
	\centering
	\caption{Truth Table for Calculation of $\hat{x}_i^{(l+1)}$.}
	\begin{tabular}{p{0.17cm}<{\centering}|p{0.17cm}<{\centering}|p{0.17cm}<{\centering}||p{6.5cm}<{\centering}}
		\Xhline{1.0pt}
		$F_1$    & $F_2$    & $F_3$   & Calculation of $\hat{x}_i^{(l+1)}$ \\
		\hline
		$0$     & $0$     & $\times$     & $\rho^{(l+1)}_{i}(\omega_{m_1})-\rho^{(l+1)}_{i}(\omega_{m_2})$ \\
		\hline
		$0$     & $1$     & $\times$     & $\rho^{(l+1)}_{i}(\omega_{m_2})-\rho^{(l+1)}_{i}(\omega_{m_1})$ \\
		\hline
		$1$     & $0$     & $0$     & $\rho^{(l+1)}_{i}(\omega_{m_2})+\rho^{(l+1)}_{i}(\omega_{m_1})+\rho^{(l+1)}_{i}(\omega_{m_2})\ll1$ \\
		\hline
		$1$     & $0$     & $1$     & $\rho^{(l+1)}_{i}(\omega_{m_2})+\rho^{(l+1)}_{i}(\omega_{m_1})+\rho^{(l+1)}_{i}(\omega_{m_1})\ll1$\\
		\hline
		$1$     & $1$     & $0$     & $-\rho^{(l+1)}_{i}(\omega_{m_2})-\rho^{(l+1)}_{i}(\omega_{m_1})-\rho^{(l+1)}_{i}(\omega_{m_2})\ll1$ \\
		\hline
		$1$     & $1$     & $1$     & $-\rho^{(l+1)}_{i}(\omega_{m_2})-\rho^{(l+1)}_{i}(\omega_{m_1})-\rho^{(l+1)}_{i}(\omega_{m_1})\ll1$ \\
		\Xhline{1.0pt}
	\end{tabular}%
	\label{tab:mean sel}%
\end{table}%

\subsubsection{Piecewise Linear Approximation}\label{s3c: PLA}
We perform piecewise linear approximation (PLA) for nonlinear operations in NNA-AMP, conducting appropriate approximation in a limited numerical interval with negligible performance loss.

\begin{figure}[htbp]
	\centering
	\hspace{-0.2cm}\subfloat[\label{fig:aprExp}]{\usepgfplotslibrary{groupplots}
\begin{tikzpicture}
\definecolor{myblued}{RGB}{0,114,189}
\definecolor{myred}{RGB}{247,25,25}
\definecolor{mypink}{RGB}{0,191,255}
\definecolor{myyellow}{RGB}{237,137,32}
\definecolor{mypurple}{RGB}{200,0,200}
\definecolor{myblues}{RGB}{77,190,238}
\definecolor{mygreen}{RGB}{0,102,80}
  \pgfplotsset{
    label style = {font=\fontsize{9pt}{7.2}\selectfont},
    tick label style = {font=\fontsize{7pt}{7.2}\selectfont}
  }

\usetikzlibrary{
    matrix,
}

\begin{axis}[
	scale = 1,
    ymode=log,
    xlabel={$E_s/N_0$ [\text{dB}]}, xlabel style={yshift=0.6em},
    ylabel={BER}, ylabel style={yshift=-0.75em},
    grid=both,
    ymajorgrids=true,
    xmajorgrids=true,
    xmin=-0.4,xmax=6.4,
    ymin=1E-04,ymax=7E-02,
    ytick={1E-04,2E-04,3E-4,4E-4,5E-4,6E-4,7E-4,8E-4,9E-4,1E-03,2E-03,3E-3,4E-3,5E-3,6E-3,7E-3,8E-3,9E-3,1E-02,2E-02,3E-2,4E-2,5E-2,6E-2,7E-2,8E-2,9E-2,1E-01},
    yticklabels={$\text{10}^{\text{-4}}$,$$,$$,$$,$$,$$,$$,$$,$$,$\text{10}^{\text{-3}}$,$$,$$,$$,$$,$$,$$,$$,$$,$\text{10}^{\text{-2}}$,$$,$$,$$,$$,$$,$$,$$,$$,$\text{10}^{\text{-1}}$},
    xtick={0, 2, 4, 6},
    xticklabels={$\text{0}$,$\text{2}$,$\text{4}$,$\text{6}$},
    grid style=dashed,
    width=0.58\columnwidth, height=0.8\columnwidth,
    thick,
    legend style={
    	nodes={scale=1, transform shape},
    	legend columns=1,
    	cells={anchor=west},
    	column sep= 1.5mm,
    	row sep= -0.25mm,
    	font=\fontsize{5.8pt}{5.2}\selectfont,
    },
    legend pos=south west,
    ]


\addplot[
    color=blue,
    mark=square*,
    line width=0.3mm,
    mark size=2.1,
]
table {
0	0.0398621875000000
1	0.0241946875000000
2	0.0133468750000000
3	0.00637625000000000
4	0.00270937500000000
5	0.000893437500000000
6	0.000239375000000000
};
\addlegendentry{NNA-AMP}

\addplot[
    color=mypink,
    mark=triangle*,
    line width=0.3mm,
    mark size=2.2,
]
table {
0	0.0408334375000000
1	0.0249784375000000
2	0.0139562500000000
3	0.00678156250000000
4	0.00291156250000000
5	0.000967500000000000
6	0.000283125000000000
};
\addlegendentry{$\hat{f}_1(-\infty, 0, \infty)$}

\addplot[
    color=red,
    mark=triangle*,
    dashed,
    every mark/.append style={solid},
    line width=0.3mm,
    mark size=2.2,
]
table {
0	0.0419734375000000
1	0.0260096875000000
2	0.0149806250000000
3	0.00755468750000000
4	0.00290812500000000
5	0.000963750000000000
6	0.000270937500000000
};
\addlegendentry{$\hat{f}_1(-4, 0, 1)$}

\addplot[
    color=mygreen,
    mark=diamond*,
    line width=0.3mm,
    mark size=2.2,
]
table {
0	0.0411562500000000
1	0.0253981250000000
2	0.0144046875000000
3	0.00712906250000000
4	0.00313187500000000
5	0.00115156250000000
6	0.000345000000000000
};
\addlegendentry{$\hat{f}_1(-2, 0, 1)$}

\addplot[
    color=myyellow,
    mark=diamond*,
    dashed,
    every mark/.append style={solid},
    line width=0.3mm,
    mark size=2.2,
]
table {
0	0.0411743750000000
1	0.0254325000000000
2	0.0144496875000000
3	0.00714875000000000
4	0.00315093750000000
5	0.00116062500000000
6	0.000348750000000000
};
\addlegendentry{$\hat{f}_1(-2, 0, 4)$}
\end{axis}

\end{tikzpicture}}\hspace{-0.4cm}
	\subfloat[\label{fig:aprDiv}]{\usepgfplotslibrary{groupplots}
\begin{tikzpicture}
\definecolor{myblued}{RGB}{0,114,189}
\definecolor{myred}{RGB}{247,25,25}
\definecolor{mypink}{RGB}{0,191,255}
\definecolor{myyellow}{RGB}{237,137,32}
\definecolor{mypurple}{RGB}{200,0,200}
\definecolor{myblues}{RGB}{77,190,238}
\definecolor{mygreen}{RGB}{0,102,80}
  \pgfplotsset{
    label style = {font=\fontsize{9pt}{7.2}\selectfont},
    tick label style = {font=\fontsize{7pt}{7.2}\selectfont}
  }

\usetikzlibrary{
    matrix,
}

\begin{axis}[
	scale = 1,
    ymode=log,
    xlabel={$E_s/N_0$ [\text{dB}]}, xlabel style={yshift=0.6em},
    ylabel={BER}, ylabel style={yshift=-0.75em},
    grid=both,
    ymajorgrids=true,
    xmajorgrids=true,
    xmin=-0.4,xmax=6.4,
    ymin=1E-04,ymax=7E-02,
    ytick={1E-04,2E-04,3E-4,4E-4,5E-4,6E-4,7E-4,8E-4,9E-4,1E-03,2E-03,3E-3,4E-3,5E-3,6E-3,7E-3,8E-3,9E-3,1E-02,2E-02,3E-2,4E-2,5E-2,6E-2,7E-2,8E-2,9E-2,1E-01},
    yticklabels={$\text{10}^{\text{-4}}$,$$,$$,$$,$$,$$,$$,$$,$$,$\text{10}^{\text{-3}}$,$$,$$,$$,$$,$$,$$,$$,$$,$\text{10}^{\text{-2}}$,$$,$$,$$,$$,$$,$$,$$,$$,$\text{10}^{\text{-1}}$},
    xtick={0, 2, 4, 6},
    xticklabels={$\text{0}$,$\text{2}$,$\text{4}$,$\text{6}$},
    grid style=dashed,
    width=0.58\columnwidth, height=0.8\columnwidth,
    thick,
    legend style={
    	nodes={scale=1, transform shape},
    	legend columns=1,
    	cells={anchor=west},
    	column sep= 1.5mm,
    	row sep= -0.25mm,
    	font=\fontsize{5.8pt}{5.2}\selectfont,
    },
    legend pos=south west,
    ]


\addplot[
    color=blue,
    mark=square*,
    line width=0.3mm,
    mark size=2.1,
]
table {
0	0.0396018750000000
1	0.0243612500000000
2	0.0133843750000000
3	0.00645687500000000
4	0.00264656250000000
5	0.000913125000000000
6	0.000255312500000000
};
\addlegendentry{NNA-AMP}

\addplot[
    color=mypink,
    mark=triangle*,
    line width=0.3mm,
    mark size=2.2,
]
table {
0	0.0404878125000000
1	0.0251965625000000
2	0.0140106250000000
3	0.00684156250000000
4	0.00283281250000000
5	0.000970625000000000
6	0.000295625000000000
};
\addlegendentry{$\hat{f}_2(-\infty, \infty, \infty)$}

\addplot[
    color=red,
    mark=triangle*,
    dashed,
    every mark/.append style={solid},
    line width=0.3mm,
    mark size=2.2,
]
table {
0	0.0409365625000000
1	0.0254825000000000
2	0.0141228125000000
3	0.00689437500000000
4	0.00283281250000000
5	0.000970625000000000
6	0.000270625000000000
};
\addlegendentry{$\hat{f}_2(\frac{1}{8}, \frac{15}{8}, 1)$}

\addplot[
    color=mygreen,
    mark=diamond*,
    dashed,
    every mark/.append style={solid},
    line width=0.3mm,
    mark size=2.2,
]
table {
0	0.0407709375000000
1	0.0253856250000000
2	0.0140509375000000
3	0.00684343750000000
4	0.00283281250000000
5	0.000970625000000000
6	0.000270625000000000
};
\addlegendentry{$\hat{f}_2(\frac{1}{8}, \frac{15}{8}, 2)$}
\end{axis}

\end{tikzpicture}}
	\caption{BER performance comparison of NNA-AMP and quantized NNA-AMP using different (a) $\hat{f}_1(\eta_{a,1}, \eta_{b,1}, N_{\text{seg},1})$ and (b) $\hat{f}_2(\eta_{a,2}, \eta_{b,2}, N_{\text{seg},2})$.}
\end{figure}
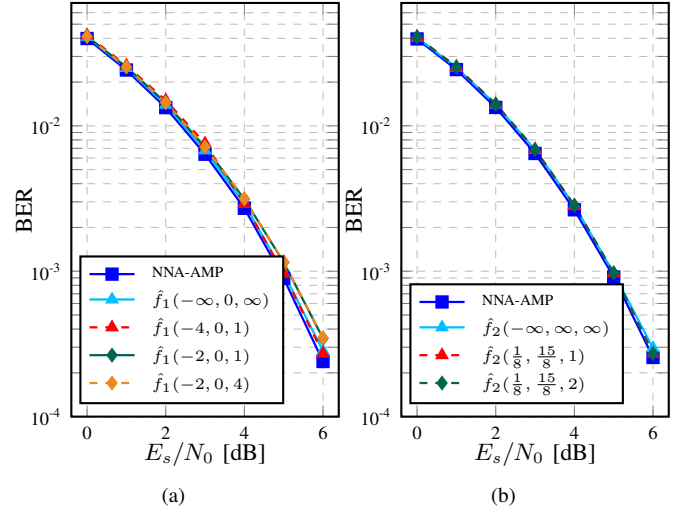

The computational complexity of NNA-AMP detector is mainly focused on the calculation of $\frac{1}{1+\exp(\widetilde{ \Delta}^{(l)}_i)}$. Considering that when the value of $\widetilde{ \Delta}^{(l)}_i (\widetilde{ \Delta}^{(l)}_i\leqslant 0)$ is small enough, $\exp(\widetilde{ \Delta}^{(l)}_i)$ can be approximately equal to $0$. We first clip $\widetilde{ \Delta}^{(l)}_i$ into range $[\eta_{a,1}, \eta_{b,1}=0]$ and then use a piecewise linear function with $N_{\text{seg},1}$ uniform segments to approximate the operations $\frac{1}{1+\exp(\cdot)}$ in range $[\eta_{a,1}, \eta_{b,1}=0]$, referred to as $\hat{f}_1(\eta_{a,1}, \eta_{b,1}, N_{\text{seg},1})$. The simulation of BER performance comparison of NNA-AMP and quantized NNA-AMP using different $\hat{f}_1(\eta_{a,1}, \eta_{b,1}, N_{\text{seg},1})$ is presented in Fig.~\ref{fig:aprExp}.

$\hat{f}_1(\eta_{a,1}, \eta_{b,1}, N_{\text{seg},1})$ and $\hat{f}_1(-\infty, 0, \infty)$ denote the quantized NNA-AMP algorithm with and without PLA function $\hat{f}_1(\eta_{a,1}, \eta_{b,1}, N_{\text{seg},1})$, respectively. As presented, when $\eta_{a,1}=-2$, no matter whether $N_{\text{seg},1}$ is $1$ or $4$, the performance is degraded compared with the quantized NNA-AMP. When $\eta_{a,1}=-4$ and $N_{\text{seg},1}=1$, the linear approximation can recover the performance of quantized NNA-AMP well. In such case, the integral part bitwidth of $\widetilde{ \Delta}^{(l)}_i$ becomes $3$ bits and a piecewise linear function $\hat{f}_1(-4, 0, 1)$ with quantized slope $0.5$ and quantized intercept $-0.125$ is chosen to approximate the nonlinear operation $\frac{1}{1+\exp(\cdot)}$.

\begin{algorithm}[htbp]\small
	\SetKwInOut{Input}{Input}\SetKwInOut{Output}{Output}
	\caption{HF-AMP Algorithm}
	\label{alg:HF-AMP}
	\Input{$\mathbf{b}, \mathbf{G}, \sigma_n^2, L,$ \\
		${r}_i^{(0)}={b}_i, \hat{x}_i^{(0)}=E_s, \bar{\xi}^{(0)}=0 (\forall i=1,...,2N_t)$.}
	\Output{$\hat{x}_i^{(L)}$ $(\forall i=1,...,2N_t)$.}
	
	\For{$l = 0, 1, ... , L-1$}
	{
		\For{$i = 1, 2, ... , 2N_t$}
		{
			$z_i^{(l)}=\hat{x}_i^{(l)}+d_i^{(l)}$;\\
			$\texttt{find }m_1\texttt{ and }m_2$;\\
			$\texttt{calculate }\widetilde{\Delta}_i^{(l)}\texttt{ according to Eq.~(\ref{Eq:Deltanew})}$;\\
			$\texttt{calculate }\rho_i^{(l)}(\omega_{m})\texttt{ in Eq.~(\ref{Eq:probabilities})}\; \texttt{with PLA}$;\\
			$\texttt{calculate } \hat{x}_i^{(l+1)} \texttt{ according to Table~\ref{tab:mean sel}}$;\\
			$d_i^{(l+1)}=b_i-\sum_jg_{i,j}\hat{x}_j^{(l+1)}$;
		}
	}
\end{algorithm}

As for $\frac{1}{\tau^{(l)}}$, according to the quantization scheme of $\tau^{(l)}$ ($1-1-3$) and $\tau^{(l)}=\sigma_n^2>0$, the range of $\tau^{(l)}$ is $[{1}/{8}, {15}/{8}]$. Therefore, we first clip $\tau^{(l)}$ into range $[\eta_{a,2}={1}/{8}, \eta_{b,2}={15}/{8}]$ and then use a piecewise linear function with $N_{\text{seg},2}$ uniform segments to approximate the division operations in range $[\eta_{a,2}={1}/{8}, \eta_{b,2}={15}/{8}]$, referred to as $\hat{f}_2(\eta_{a,2}, \eta_{b,2}, N_{\text{seg},2})$. The simulation of BER performance with different $\hat{f}_2(\eta_{a,2}, \eta_{b,2}, N_{\text{seg},2})$ is presented in Fig.~\ref{fig:aprDiv}. $\hat{f}_2(\eta_{a,2}, \eta_{b,2}, N_{\text{seg},2})$ with $N_{\text{seg},2}=1$ is enough to recover the performance of the quantized NNA-AMP. Therefore, a piecewise linear function $\hat{f}_2({1}/{8}, {15}/{8}, 1)$ with quantized slope $-4.25$ and quantized intercept $8.5$ is chosen to approximate the division operation, ensuring both ease of hardware implementation and BER performance.
\begin{Rem}
	Unlike the conventional implementation of nonlinear operations where all possible values are directly stored in read-only memory (ROM) for look-up tables (LUTs), using PLA only needs to store the interval, slope, and intercept. In such case, the number of values to be stored is notably reduced. The storage overhead can be effectively alleviated. Compared with traditional generation methods, e.g., coordinate rotation digital computer (CORDIC) algorithm, the processing latency is considerably shortened to the processing time of a multiplier and an adder.
\end{Rem}

The NNA-AMP with NC and strength reduction, referred to as HF-AMP, is summarized in Alg.~\ref{alg:HF-AMP}.
Meanwhile, the quantization scheme of HF-AMP according to Table~\ref{tab:quan} is listed in Table~\ref{tab:quan-HF}.
\begin{table}[htbp]
	\tabcolsep 1mm
	\renewcommand{\arraystretch}{1.2}
	\centering
	\footnotesize
	\caption{Quantization Scheme of HF-AMP Based on AHPQ.}
	\label{tab:quan-HF}
	\begin{tabular}{c||p{0.28cm}<{\centering}|p{0.28cm}<{\centering}|p{0.28cm}<{\centering}|p{0.28cm}<{\centering}|p{0.28cm}<{\centering}|p{0.28cm}<{\centering}|p{0.28cm}<{\centering}|p{0.28cm}<{\centering}|p{0.28cm}<{\centering}|p{0.28cm}<{\centering}|p{0.28cm}<{\centering}|p{0.28cm}<{\centering}|p{0.28cm}<{\centering}|p{0.28cm}<{\centering}}
		\Xhline{1.0pt}
		\multicolumn{1}{c||}{$k$} & 1 & 2 & 3 & 4 & 5 & 6 & 13 & 14 & 15 & 16 & 17 & 18 & 20 & 21 \\ \hline
		\multicolumn{1}{c||}{Integal Part} & \multicolumn{1}{c|}{3} & \multicolumn{1}{c|}{2} & \multicolumn{1}{c|}{1} & \multicolumn{1}{c|}{1} & \multicolumn{1}{c|}{2} & \multicolumn{1}{c|}{2} & \multicolumn{1}{c|}{1}  & \multicolumn{1}{c|}{4}  & \multicolumn{1}{c|}{3}  & \multicolumn{1}{c|}{4}  & \multicolumn{1}{c|}{3}  & \multicolumn{1}{c|}{3}  & \multicolumn{1}{c|}{6}  & 3 \\ \hline
		\multicolumn{1}{c||}{Fractional Part} &  \multicolumn{1}{c|}{6} & \multicolumn{1}{c|}{7} & \multicolumn{1}{c|}{3} & \multicolumn{1}{c|}{3} & \multicolumn{1}{c|}{3} & \multicolumn{1}{c|}{2}  & \multicolumn{1}{c|}{3}  & \multicolumn{1}{c|}{1}  & \multicolumn{1}{c|}{4}  & \multicolumn{1}{c|}{4}  & \multicolumn{1}{c|}{6}  & \multicolumn{1}{c|}{4}  & \multicolumn{1}{c|}{1}  & 1  \\ \Xhline{1.0pt}
	\end{tabular}
\end{table}

\subsection{Overall VLSI Architecture}

\begin{figure}[htbp]
	\centering
	\includegraphics[width=1\columnwidth]{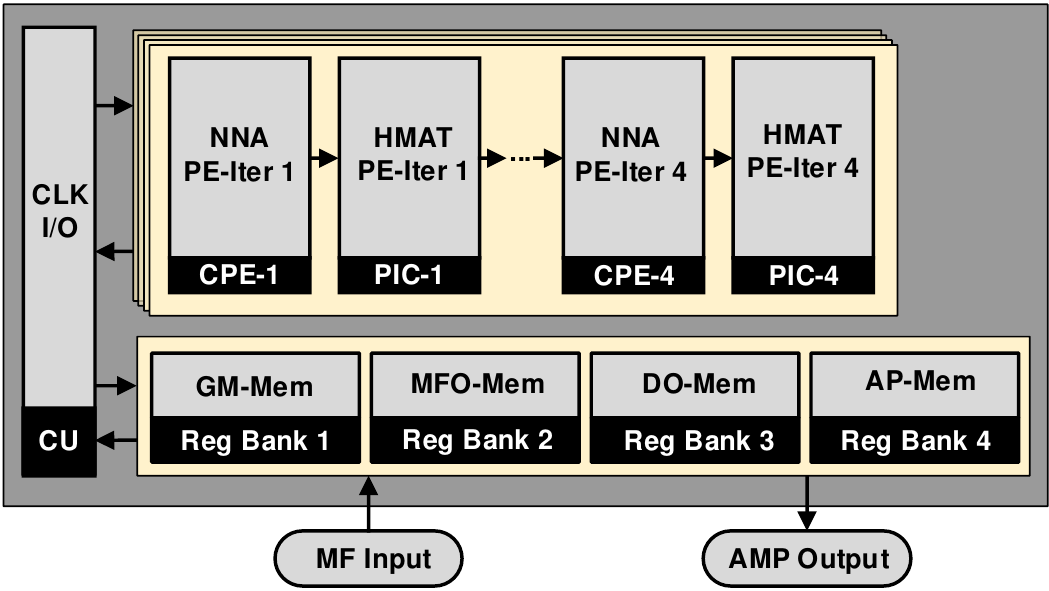}
	\caption{Hardware implementation architecture for HF-AMP detector. HMAT: hybrid-precision multiplier and adder tree; MF: matched filter; GM: Gram matrix; MFO: matched filter output; DO: detector output; AP: auxiliary parameters; Reg: regester; Mem: memory.}
	\label{fig:NNA_AMP}
\end{figure}

The system configuration of HF-AMP is the same as Section~\ref{sec: Application Examples}. With high-throughput and low-complexity design, the overall architecture is demonstrated in Fig.~\ref{fig:NNA_AMP}, composed of two main processing elements (PEs) named \texttt{constellation processing element} (\texttt{CPE}) and \texttt{parallel interference cancellation} (\texttt{PIC}) in each pipeline iteration, a \texttt{control unit} (\texttt{CU}) for clock and input/output (I/O) control, and \texttt{register banks} for storing the I/O and auxiliary data.

\subsubsection{\texttt{CPE module}}\label{s3c:NNA}
\begin{figure}[htbp]
	\centering
	\includegraphics[width=1\columnwidth]{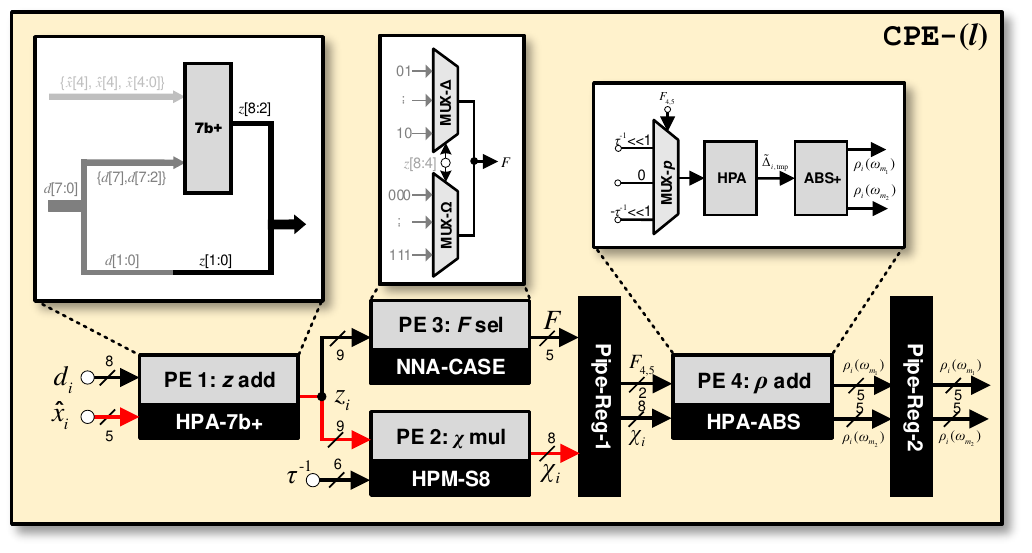}
	\caption{Pipeline architecture for \texttt{CPE module}. sel: select; mul: multiply.}
	\label{fig:NNA_PE}
\end{figure}

The branch selection signal $\{F_1, F_2, F_3, F_4, F_5\}$ and the probability $\rho^{(l+1)}_i(\omega_{m_1})$ and $\rho^{(l+1)}_i(\omega_{m_2})$ are determined in \texttt{CPE module} illustrated in Fig.~\ref{fig:NNA_PE}, where $d[w_a:w_b]$ ($\hat{x}[w_a:w_b]$, $z[w_a:w_b]$) denotes the $w_a$-th bit to $w_b$-th bit of variable $d_i^{(l)}$ ($\hat{x}_i^{(l)}$, $z_i^{(l)}$).

In this unit, variable $z_i^{(l)}$ in Line $3$ of Alg.~\ref{alg:HF-AMP} is calculated by hybrid-precision adder (HPA) of \texttt{HPA-7b+ module}. Note that the superscript ${(l)}$ of variables is omitted for convenience in Figs.~\ref{fig:NNA_PE}, \ref{fig:PIC}, \ref{fig:SAA}, and \ref{fig:MACT}. Since the quantization schemes of input variables $d_i^{(l)}$, input variables $\hat{x}_i^{(l)}$, and output variables $z_i^{(l)}$ are $1-3-4$, $1-2-2$, and $1-4-4$, respectively, we consider to implement the two's complement addition without overflow and use adders with smallest bitwidths in \texttt{HPA-7b+ module}. In detail, the HPA is designed as a $7$-bit half-adder. The operand $\hat{x}_i^{(l)}$ ($\hat{x}[4:0]$) and part of operand $d_i^{(l)}$ ($d[7:2]$) are firstly extended to $7$ bits. After the $7$-bit half-adder, the output is assigned to $z[8:2]$. The rest bits of $z_i^{(l)}$ ($z[1:0]$) is directly equal to $d[1:0]$.

For calculating $\chi_i^{(l)}$ in Eq.~(\ref{Eq:Deltanew}), a hybrid-precision multiplier (HPM) is employed in \texttt{HPM-S8 module}. The two input operands of HPM in \texttt{HPM-S8 module} are $9$ bits ($1-4-4$) and $6$ bits ($1-4-1$), respectively, and $8$ bits ($1-6-1$) are taken as the output.

To obtain $m_1$ and $m_2$, \texttt{NNA-CASE module}, including two multiplexers \texttt{MUX-$\Delta$} and \texttt{MUX-$\Omega$}, first extracts the integral part of the auxiliary variable $z^{(l)}_i$ and then selects the corresponding flags. \texttt{MUX-$\Delta$} determines $F_4, F_5$ and \texttt{MUX-$\Omega$} determines $F_1, F_2, F_3$. After that, $F_4, F_5$ are utilized to determine how to calculate $\widetilde{\Delta}_{i,\mathrm{tmp}}^{(l)}$ as follows,
\begin{equation}
	\widetilde{\Delta}_{i,\mathrm{tmp}}^{(l)}=\left\{
	\begin{aligned}
		&\chi_i^{(l)} + (\frac{1}{\tau^{(l)}}\ll 1), &\{F_4,F_5\}=01\\
		&\chi_i^{(l)} - (\frac{1}{\tau^{(l)}}\ll 1), &\{F_4,F_5\}=10\\
		&\chi_i^{(l)}, &\{F_4,F_5\}=11\\
	\end{aligned}
	\right..
\end{equation}

As discussed in Section~\ref{s3c: PLA}, the quantized slope and intercept of the linear interpolation function $\hat{f}_1$ are $0.5$ and $-0.125$, respectively, which is convenient for hardware implementation by performing shift operations. Then $\rho^{(l+1)}_i(\omega_{m_1})$ and $\rho^{(l+1)}_i(\omega_{m_2})$ are calculated separately in \texttt{ABS+ module} as follows,
\begin{small}
\begin{equation}
	\left\{
	\begin{aligned}
		&\rho^{(l+1)}_i(\omega_{m_1})= \frac{1}{2}\cdot \widetilde{ \Delta}^{(l)}_i - \frac{1}{8} = -\left|\widetilde{\Delta}_{i,\mathrm{tmp}}^{(l)}\right|- (0.001)_2\\
		&\rho^{(l+1)}_i(\omega_{m_2})= -\frac{1}{2}\cdot \widetilde{ \Delta}^{(l)}_i + \frac{7}{8} = \left|\widetilde{\Delta}_{i,\mathrm{tmp}}^{(l)}\right|+ (0.111)_2\\
	\end{aligned}
	\right..
\end{equation}
\end{small}
The adders used in \texttt{HPA module} and \texttt{ABS+ module} are similar to the HPA in \texttt{HPA-7b+ module}.

\subsubsection{\texttt{PIC module}}
\begin{figure}[htbp]
	\centering
	\includegraphics[width=0.75\columnwidth]{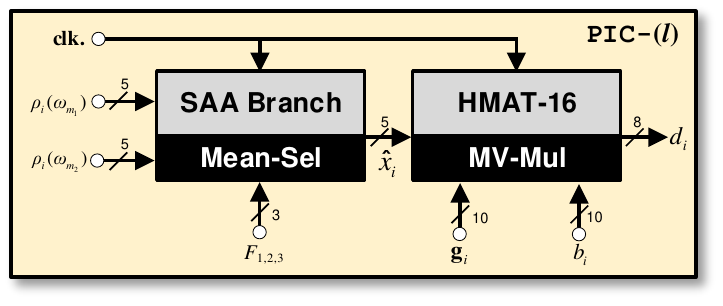}
	\caption{Hardware architecture for \texttt{PIC module}. HMAT: hybrid-precision multiplier and adder tree.}
	\label{fig:PIC}
\end{figure}

As presented in Fig.~\ref{fig:PIC}, \texttt{PIC module} consists of \texttt{Mean-Sel module} and \texttt{MV-Mul module}, which complete the calculation of $\hat{x}_i^{(l+1)}$ (Line 7 in Alg.~\ref{alg:HF-AMP}) and the calculation of $d_i^{(l+1)}$ (Line 8 in Alg.~\ref{alg:HF-AMP}), respectively.

\begin{figure}[htbp]
	\centering
	\includegraphics[width=0.65\columnwidth]{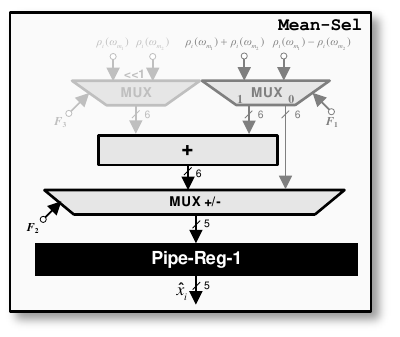}
	\caption{Hardware architecture for \texttt{Mean-Sel module}.}
	\label{fig:SAA}
\end{figure}

\begin{figure}[htbp]
	\centering
	\includegraphics[width=0.9\columnwidth]{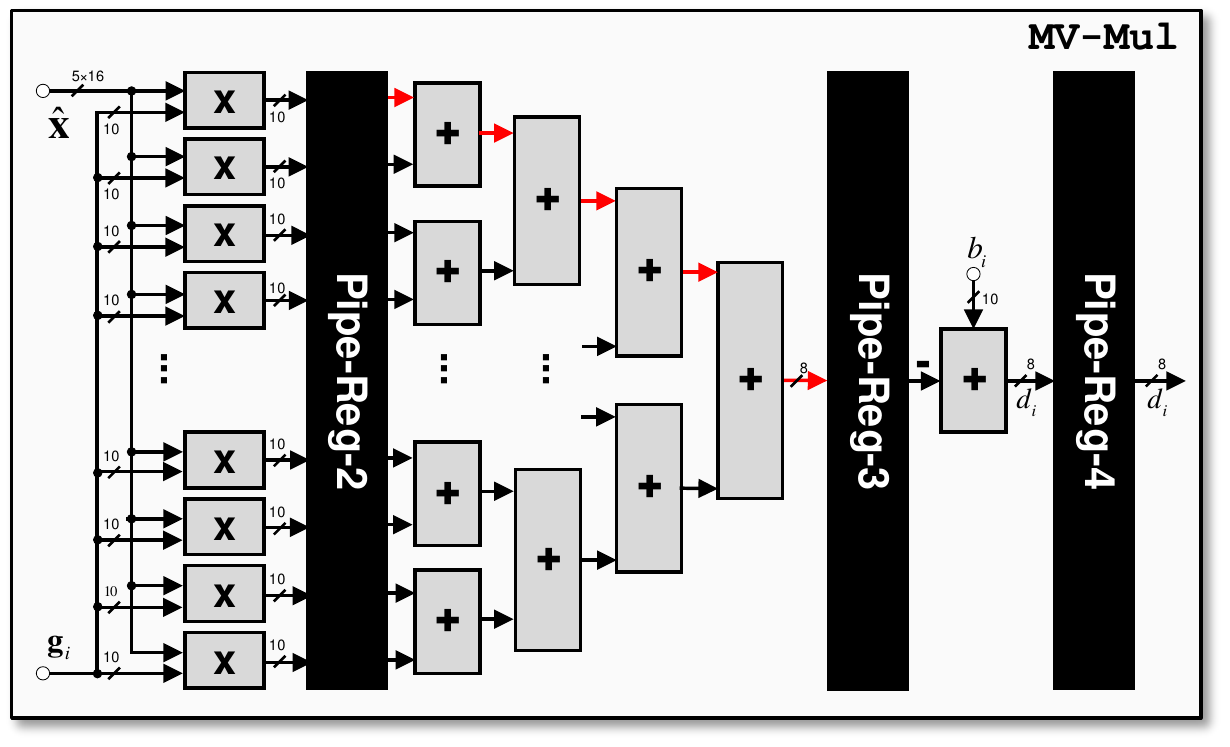}
	\caption{Hardware architecture for \texttt{MV-Mul module}.}
	\label{fig:MACT}
\end{figure}

\texttt{Mean-Sel module} based on Table~\ref{tab:mean sel} are illustrated in Fig.~\ref{fig:SAA}. Via the flags $\{F_1, F_2, F_3\}$, a two-stage selector determines selective branches of $\hat{x}_i^{(l+1)}$.
In \texttt{MV-Mul module}, to implement the multiplication of the Gram matrix $\mathbf{G}$ and the vector $\hat{\mathbf{x}}^{(l+1)}=[\hat{x}_1^{(l+1)},...,\hat{x}_{2N_t}^{(l+1)}]^{\mathsf{T}}$ with low processing latency, the matrix-vector multiplication is transformed into $2N_t$ groups of vectors' inner-product of $\mathbf{g}_i=[g_{i,1},...,g_{i,2N_t}]^{\mathsf{T}}$ and $\hat{\mathbf{x}}^{(l+1)}$, and each group is executed by HPMs and a four-stage HPA tree as shown in Fig.~\ref{fig:MACT}. Another adder is used to combine the output of adder tree and $b_i$ to obtain $d_i^{(l+1)}$.

\subsection{Time Analysis for Full-pipeline Architecture}
\begin{table*}[htpb]
	\centering
	\renewcommand{\arraystretch}{1.2}
	\caption{Detailed Data Processing Order Based on the Full-Parallel Pipeline Design}
	\begin{tabular}{c||c|c|c|c|c|c|c|c|c|c}
		\Xhline{1.0pt}
		\multirow{1}[3]{*}{\textbf{Stage}}  & \multicolumn{6}{c|}{$1$-st iteration}    &  $\cdots$  & \multicolumn{3}{c}{$4$-th iteration} \\
		\cline{2-11}     & CPE-$1$:R-$1^\text{a}$& CPE-$1$:R-$2^\text{a}$& PIC-$1$:R-$1^\text{a}$& PIC-$1$:R-$2^\text{a}$& PIC-$1$:R-$3^\text{a}$& PIC-$1$:R-$4^\text{a}$& $\cdots$ & $\cdots$     & PIC-$4$:R-$3^\text{a}$& PIC-$4$:R-$4$(OUT-R)$^\text{a}$ \\
		\hline
		\hline
		$\kappa$     & $F_\kappa$    &$F_{\kappa-1} $ &$F_{\kappa-2} $ &$F_{\kappa-3}$  &$F_{\kappa-4} $ &$F_{\kappa-5} $ &$\cdots$ & $\cdots$     &$F_{\kappa-22}$ &$F_{\kappa-23}$ \\
		\hline
		$\kappa+1$   & $F_{\kappa+1}$  &$F_{\kappa} $ &$F_{\kappa-1 }$ &$F_{\kappa-2 }$ &$F_{\kappa-3 }$ &$F_{\kappa-4 }$ &$\cdots$ & $\cdots$     &$F_{\kappa-21 }$&$F_{\kappa-22 }$\\
		\hline
		$\kappa+2$   &$F_{\kappa+2}$  &$F_{\kappa+1 } $&$F_{\kappa} $  &$F_{\kappa-1 } $&$F_{\kappa-2 }$ &$F_{\kappa-3 }$ &$\cdots$ & $\cdots$     &$F_{\kappa-20 }$&$F_{\kappa-21 }$\\
		\hline
		$\cdots$     & $\cdots$     & $\cdots$     & $\cdots$     & $\cdots$     & $\cdots$     & $\cdots$     & $\cdots$     & $\cdots$     & $\cdots$          & $\cdots$ \\
		\hline
		$\kappa+23$  &$F_{\kappa+23 }$&$F_{\kappa+22 }$&$F_{\kappa+21 }$&$F_{\kappa+20}$&$F_{\kappa+19}$&$F_{\kappa+18}$&$\cdots$& $\cdots$     &$F_{\kappa+1} $  &$F_{\kappa }$\\
		\Xhline{1.0pt}
	\end{tabular}%
	\label{tab:time}%
	\begin{tablenotes}
		\item[1] \hspace{1mm} \scriptsize{$^\text{a}$  CPE-$l$:R-$\iota(\iota=1,2)$ and PIC-$l$:R-$\iota(\iota=1,2,3,4)$ represent the pipeline register groups in \texttt{CPE module} and \texttt{PIC module} at the $l$-th iteration, respectively. $F_\kappa$ in CPE-$l$:R-$\iota$ (or PIC-$l$:R-$\iota$) means the data of $\kappa$-th frame is processed and stored to the CPE-$l$:R-$\iota$ (or PIC-$l$:R-$\iota$) during the $(\kappa-1, \kappa]$ clock.}
	\end{tablenotes}
\end{table*}

Table~\ref{tab:time} illustrates the time schedule of fine-granularity HF-AMP. Four iterations are fully unfolded to improve the throughput of the detector, and each iteration consists of a \texttt{CPE module} and a \texttt{PIC module}. In the non-iterative \texttt{CPE module}, the overall path is divided into two parts by two register groups in Fig.~\ref{fig:NNA_PE}, and the processing time of each part is less than or equal to the processing time of HPM and HPA, $t_\text{HMA}$, to ensure the shortest critical path of the module. In the non-iterative \texttt{PIC module}, the overall path is divided into four parts by four register groups as shown in Figs.~\ref{fig:SAA} and \ref{fig:MACT}, and each part is less than or equal to $t_\text{HAT}$ of the HPA tree. The main system is composed of $\mathcal{P}=24$-level full pipeline. Each frame is calculated within $24$ clock cycles. At each moment, $24$ frames are processed simultaneously, which achieves the maximum utilization of pipeline architecture.

\begin{Rem}
	For the trade-off between system throughput and area power consumption, the granularity of the pipeline needs to be determined. Different quantization schemes affect the processing delay of the computational units, which affects the critical path of the system. Therefore, we choose the maximum value $\max\{t_\text{HMA},t_\text{HAT}\}$ of the critical paths in the two main modules as the critical path of the system.
\end{Rem}

\section{Hardware Implementation}\label{sec: Hardware Verification}
\subsection{Experimental Setup}\label{s2c: ES}
The HF-AMP in Fig.~\ref{fig:NNA_AMP} is implemented in SMIC $65$ nm LL $1$P$9$M CMOS technology based on the design parameters in Section~\ref{sec: HF AMP and VLSI}. The HF-AMP detector is synthesized on Synopsys\textsuperscript{\textregistered} Design Compile (DC) and the results are placed and routed using the Synopsys\textsuperscript{\textregistered} IC Compiler. The annotated toggle rate of the gate-level netlist is converted into switching activity interchange format (SAIF) for Prime-Time PX to measure the time-based chip-only power dissipation \cite{xu2020reconfigurable}.

\subsection{Implementation Results}
\subsubsection{ASIC results}
Denote the clock frequency as $f_\mathrm{clk}$. The throughput of the HF-AMP detector can be defined as:
\begin{equation}
\text{T/P} \mathrm{[Gb/s]} = \frac{\log_2(Q)\cdot N_t \cdot \mathcal{P}}{T_\mathrm{cyc}} \cdot f_\mathrm{clk},
\end{equation}
where $T_\mathrm{cyc}$ is the average detecting cycles.

\begin{figure}[htpb]
	\centering
	\subfloat[]{\includegraphics[width=0.52\columnwidth]{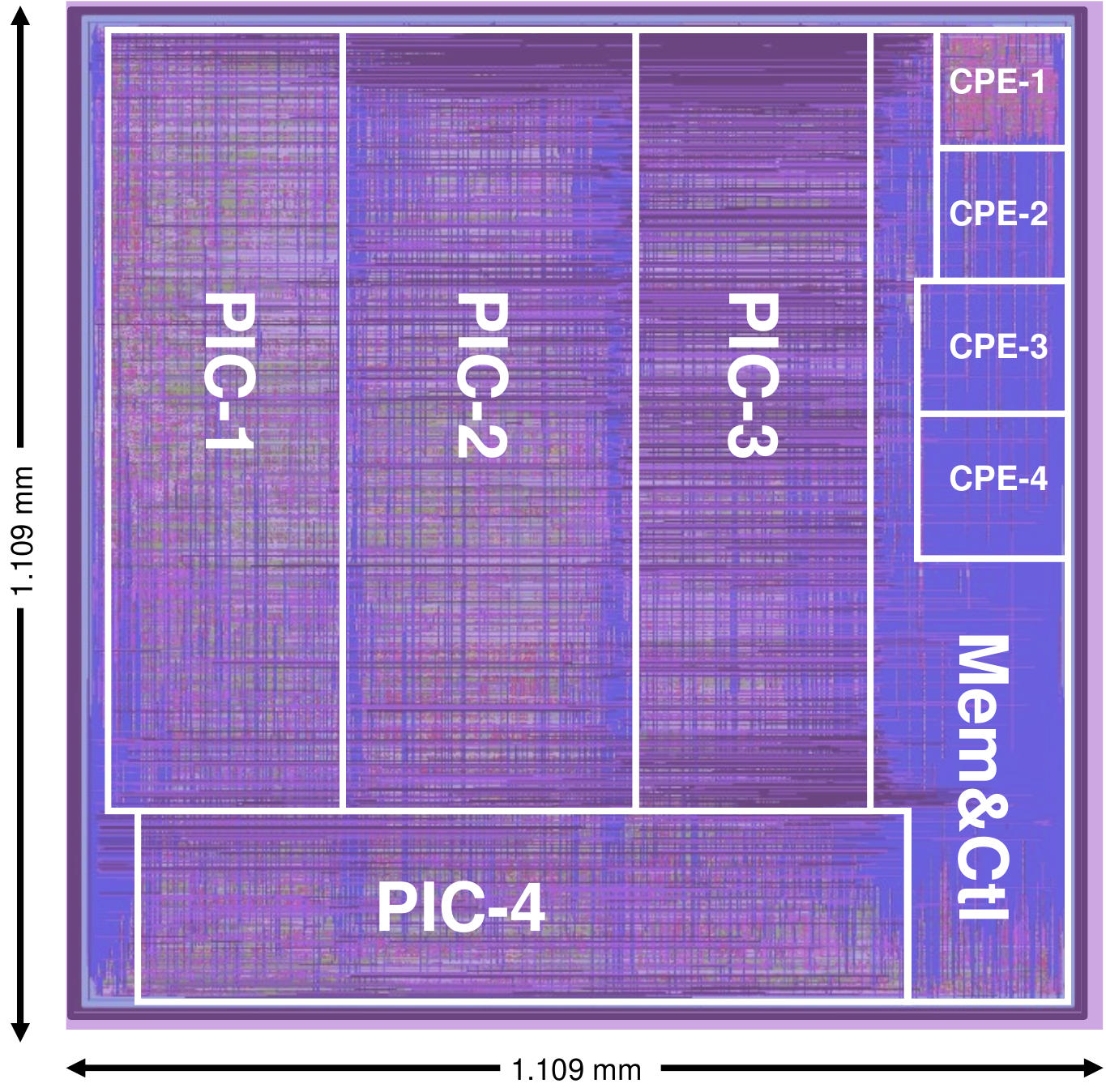}
		\label{subfig:layout}}
	\hfil
	\subfloat[]{\includegraphics[width=0.72\columnwidth]{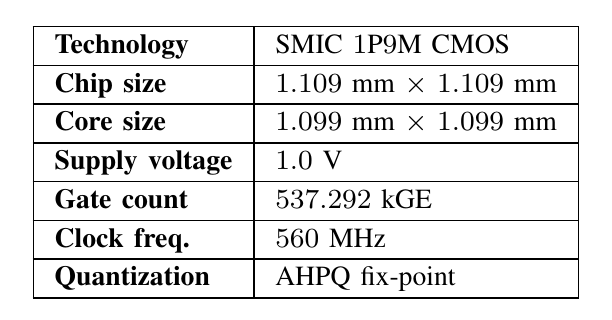}
		\label{subfig:chip_specification}}
	\hfil
	\subfloat[]{\includegraphics[width=0.43\columnwidth]{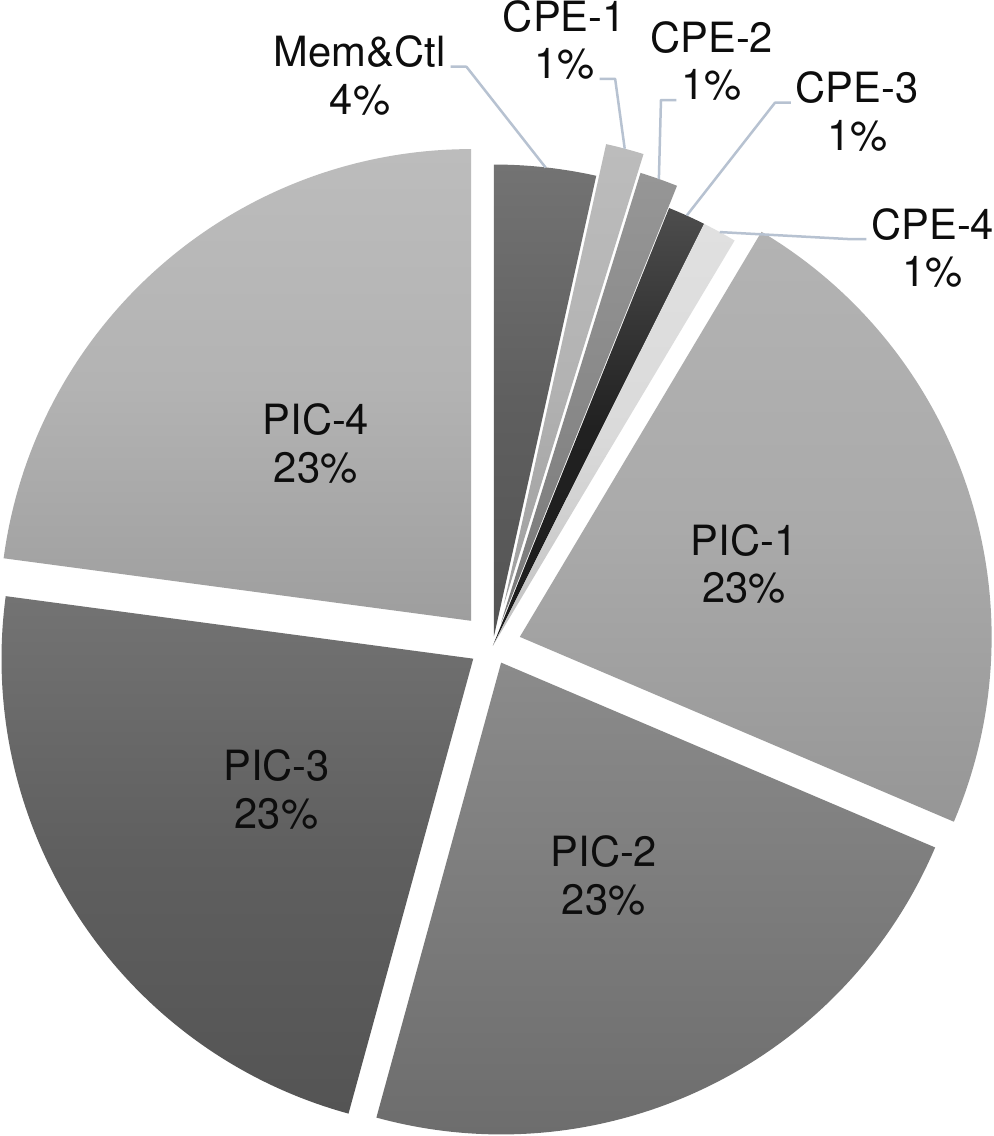}
		\label{subfig:area_breakdown}}
	\caption{(a) Layout photo, (b) chip specification, and (c) area breakdown of the HF-AMP in SMIC 65nm technology.}
	\label{fig:layout}
\end{figure}

This work presents the \textit{first} hardware design of the AHPQ-assisted HF-AMP detector. Fig.~\ref{fig:layout} illustrates the layout photo, detailed summary of the chip specification, and area breakdown of the HF-AMP detector, where freq. denotes frequency. Under SMIC $65$ nm CMOS technology, the whole chip consists of $537.292$k logic gates integrated within $1.208$ mm$^2$ and dissipates $142.05$ mW with $1.0$ V supply and $560$ MHz clock. $128\times8$ massive MIMO detector achieves a peak throughput of $17.92$ Gb/s.

\subsubsection{Comparison between AHPQ and UQ}\label{subsubsec: Comparison}
We implement the HF-AMP detector quantized by AHPQ and UQ separately with ASIC tools mentioned above and Xilinx Virtex-7 xc7v2000tfhg 1761-1 FPGA. The results are listed in Table~\ref{tab:ASIC}, where syn. and eff. denote synthesis and efficiency, respectively. In the same ASIC library and environmental parameters, the chip core area of HF-AMP using AHPQ is only $57.5\%$ of that of HF-AMP using UQ. Benefit from smaller quantization bitwidth, the critical path of HF-AMP using AHPQ shrinks to only $64.3\%$ compared with HF-AMP using UQ. HF-AMP using AHPQ can achieve nearly $2.96 \times$ ASIC area efficiency. The bitwidth reduction reduces the energy consumption of HF-AMP using UQ by about $80.1\%$. Meanwhile, HF-AMP using AHPQ can speed up the ASIC synthesis process by about $23.4\%$.
For field-programmable semi-custom devices, the resource consumption of LUT slices, DSP units, and FF slices is reduced by $62.7\%$, $94.1\%$, and $74.9\%$, respectively.

\begin{table}[htpb]
	\renewcommand{\arraystretch}{1.2}
	\caption{ASIC and FPGA Implementation Results of HF-AMP with AHPQ and HF-AMP with UQ.}
	\tabcolsep 2mm
	\begin{center}
		\begin{tabular}{l |c c}
			\specialrule{0.12em}{0.5pt}{0.5pt}
			\textbf{Comparison Item}& HF-AMP (AHPQ) & HF-AMP (UQ) \\
			\hline
			\hline
			\rowcolor{mygray}\textbf{ASIC area [$\text{mm}^2$]}  & $1.208$ & $2.100$\\
			
			\textbf{ASIC T/P [Gb/s]} & $17.92$ & $10.56$\\
			
			\rowcolor{mygray}\textbf{ASIC gate count$^{\text{\ding{172}}}$ [kGE]}   & $537.292$ & $1,202.768$\\
			
			\textbf{ASIC syn. cost [HH-mm-ss] $^{\text{\ding{173}}}$ } & $00$-$58$-$10$ & $01$-$29$-$01$\\
			
			\rowcolor{mygray}\textbf{ASIC critical path [ns]} & $1.80 (t_\text{HAT})$ & $2.80 (t_\text{HMA})$\\
			
			\textbf{ASIC energy eff. $^{\text{\ding{174}}}$ [pJ/b]}  &  $7.93$ & $9.83$\\
			
			\rowcolor{mygray}\textbf{ASIC area eff. $^{\text{\ding{175}}}$ [Gb/s/mm$^2$]} &$14.93$ & $5.03$\\
			\hline
			\hline
			
			\textbf{FPGA LUT slices} & $12,985$ & $20,703$\\
			
			\rowcolor{mygray}\textbf{FPGA DSP units} & $1,024$ & $1,088$\\
			
			\textbf{FPGA FF slices} & $6,016$ & $8,032$\\
			\specialrule{0.12em}{0.5pt}{0.5pt}
		\end{tabular}
		\begin{tablenotes}
			\item[1] \scriptsize{$^{\text{\ding{172}}}$Estimated by 2-input NAND gate in ASIC.}\ \ \ \  \scriptsize{$^{\text{\ding{173}}}$Time consuming for DC synthesis.}
			\item[3]  \scriptsize{$^{\text{\ding{174}}}$ASIC energy eff. is defined as energy consumption per bit.}
			\item[4]  \scriptsize{$^{\text{\ding{175}}}$ASIC area eff. is defined as TAR.}
		\end{tablenotes}
	\end{center}
	\label{tab:ASIC}
\end{table}%

\subsubsection{Comparison to Prior Arts}
To make a fair comparison, the SOA detectors are normalized to the same feature size and threshold voltage as
\begin{equation}
f_\mathrm{clk} \propto \zeta , A \propto \frac{1}{\zeta^2}, E \propto \frac{\upsilon^2}{\zeta},
\end{equation}
where $A$ and $E$ are core area and energy cost, respectively. $\zeta$ and $\upsilon$ denote the scaling factor and voltage factor \cite{peng2017ASIC}. Furthermore, the system performance should be normalized to the same antenna configuration and modulation mode. Consider the number of transmitting antennas is normalized to $8$, the throughput, chip area, and power consumption are all scaled by $\frac{8}{N_t}\frac{log_2 N_t}{log_2 8}$\cite{tan2019improving}. For the normalization of the modulation mode, the area scaling is determined by the computational complexity for soft-output MPD detectors \cite{chen2018integrated}; for hard-output MPD detectors and minimum mean-squared error (MMSE) detectors, the area scaling is almost independent of the modulation mode.

\begin{table*}[htp]
	\renewcommand{\arraystretch}{1.2}
	\caption{ASIC Implementation Results \& Comparison for MIMO Detectors.}
	\begin{center}
		\begin{tabular}{l||m{2.5cm}<{\centering}|m{1.8cm}<{\centering}|m{1.8cm}<{\centering}|m{2cm}<{\centering}|m{1.8cm}<{\centering}|m{1.8cm}<{\centering}}
			\Xhline{1pt}
			\multirow{2}{*}{\textbf{Detector}} &  \multirow{2}[1]{*}{This Work}  & X. Tan~\cite{tan2019improving}  & G. Peng~\cite{peng2017ASIC}  & Y.-T. Chen~\cite{chen2018integrated} &  W. Tang~\cite{tang20210} & C. Jeon~\cite{jeon2019354}\\
			
			&    & [TVT'20] & [TCASI'18]  & [TCASI'19]   & [JSSC'21] & [LSSC'19]\\
			\hline
			\rowcolor{mygray}\textbf{Algorithm} & HF-AMP (AHPQ)$^{\text{\ding{172}}}$ & MPD-DNN$^{\text{\ding{172}}}$ & MMSE & Soft-MPD$^{\text{\ding{172}}}$   & MPD$^{\text{\ding{172}}}$ &LAMA$^{\text{\ding{172}}}$\\
			\textbf{MIMO size}& $128\times8$  &$128\times8$   &$128\times8$  &$128\times8$   &$128\times32$ &$128\times32$ \\
			\rowcolor{mygray}\textbf{Modulation} & 16-QAM  & 16-QAM & 64-QAM & QPSK&256-QAM&256-QAM\\
			\textbf{Technology (SV)} & SMIC 65 (1.0)  & SMIC 65 (1.0) & SMIC 65 (1.0) & TSMC 40 (0.9) & TSMC 40 (0.9)& TSMC 28 (0.9)\\
			\hline
			\rowcolor{mygray}\textbf{Area [$\text{mm}^2$]} & $1.208$  & $0.801$ & $2.570$ & $0.076$ & $0.580$& $0.370$\\
			\textbf{Max. freq. [MHz]} & $560$ &  $340$ & $680$ & $500$ & $425$& $400$ \\
			\rowcolor{mygray}\textbf{T/P [Gb/s]} & $17.92$  & $0.18$ & $1.02$ & $8.00$& $2.76$& $0.35$\\
			\textbf{Power [mW]} & $142.05$ & N/A & $65.00$ & $77.93$ & $220.6$& $151.0$\\
			\rowcolor{mygray}\textbf{Energy [pJ/b]} & $7.93$  & N/A & $63.72$  & $9.74$& $79.8$& $426$\\
			\hline
			\textbf{Norm. T/P$^{\text{\ding{173}}}$ [Gb/s]} &$17.92$ & $0.18$ & $1.02$  & $9.85$& $0.35$& $0.05$\\
			\rowcolor{mygray}\textbf{Norm. TAR$^{\text{\ding{173}}}$ [Gb/s/$\text{mm}^2$]} & $14.93$  & $0.23$ & $0.40$ & $16.35$&  $1.66$&  $0.33$\\
			\textbf{Norm. energy$^{\text{\ding{173}}}$ [pJ/b]} & $7.93$  & N/A  & $63.72$ & $15.88$& $521.13$& $2,781.20$ \\
			\Xhline{1pt}
		\end{tabular}
		\begin{tablenotes}
			\item[1] \hspace{1mm} \scriptsize{$^{\text{\ding{172}}}  $Preprocessing unit for the matching filter $\mathbf{b}$ and $\mathbf{g}$ are not included.}\ \ \ \ \scriptsize{$^{\text{\ding{173}}} $Scaled to $N_t=8$, $16$-QAM and SMIC $65$ nm with $1.0$ V supply.}
		\end{tablenotes}
	\end{center}
	\label{tab:hardware_comparision}
\end{table*}%
Table~\ref{tab:hardware_comparision} summarizes the comparison results between our proposed HF-AMP using AHPQ and the SOA detectors, including MPD-DNN \cite{tan2019improving}, MMSE \cite{peng2017ASIC}, soft-MPD \cite{chen2018integrated}, and MPD \cite{tang20210}, where SV, max. freq., and norm. mean supply voltage, maximum frequency, and normalized. The proposed HF-AMP using AHPQ brings considerable improvements in both area efficiency and energy efficiency, achieving a $37.3\times$ higher Norm. TAR and $8.04\times$ less bit energy compared with traditional linear detectors MMSE \cite{peng2017ASIC}. Meanwhile, it has been demonstrated in \cite{jeon2018optimal} that AMP can achieve near-optimal performance in massive MIMO limit.

Introduction of the efficient AHPQ allows the operation nodes to be greatly simplified, ensuring that the iterative signal detection process can be fully unfolded while meeting the energy and area efficiency requirements. The minimal pipeline granularity ensures the highest throughput for our fully-unfolded pipeline architecture. Compared to MPD detector \cite{tang20210}, we achieve an $8.99\times$ improvement in Norm. TAR and reduce $65.72\times$ energy consumption. Notably, compared to the DNN-assisted MPD detector \cite{tan2019improving}, the hardware utilization of different modules in our implementation is maximized due to the fully-unfolded pipeline architecture without the inter-module data waiting, which results in a $64.91\times$ TAR boost. We also acquire $2.00\times$ energy saving of soft-MPD proposed by \cite{chen2018integrated} with a comparable Norm. TAR. Compared with \cite{jeon2019354}, our implementation presents obvious advantages in both Norm. TAR and Norm. energy. Therefore, our proposed HF-AMP benefitting from AHPQ and fully-unfolded pipeline architecture provides superior area and energy efficiency.

\section{Conclusion}\label{sec: Conclusion}
In this paper, an AHPQ for MIMO detectors, including PDF-based IPQ and DRL-based FPQ, is proposed. Its application on the NNA-AMP detector is analyzed, showing that AHPQ results in much lower quantization bitwidth than UQ. The hardware implementations of HF-AMP are presented to demonstrate the advantage of AHPQ. With a fully-unfolded pipeline architecture designed under SMIC 65nm CMOS technology, the HF-AMP detector reaches a high throughput of $17.92$ Gb/s with $142.05$ mW for $128\times 8$ $16$-QAM MIMO. Compared to the SOA, HF-AMP using AHPQ enjoys advantages in both area and energy efficiency.

It is worth mentioning that our AHPQ scheme can also be applied to other digital signal processing modules, not just MIMO detectors. Due to the page limit, those applications are not discussed in this paper but in our future papers.

\bibliographystyle{IEEEtran}
\bibliography{Ref/IEEEabrv,Ref/mybib}

\end{document}